\documentstyle[tighten,aps,twocolumn,epsfig,amsfonts]{revtex}

\newcommand{\R}{{\rm I\!R}}
\newcommand{\plus}{{\!+\!}}
\newcommand{\minus}{{\!-\!}}

\newcommand{\vsp}{\vspace*{3mm}}
\newcommand{\pprime}{{\prime\prime}}
\newcommand{\bra}{\langle}
\newcommand{\ket}{\rangle}
\newcommand{\order}{{\cal O}}
\newcommand{\bOmega}{\bbox{\Omega}}
\newcommand{\bxi}{\bbox{\xi}}
\newcommand{\bomega}{\bbox{\omega}}
\newcommand{\bpsi}{\bbox{\psi}}
\newcommand{\hC}{\widehat{C}}

\newcommand{\hL}{\widehat{L}}
\newcommand{\hK}{\widehat{K}}
\newcommand{\bR}{\bbox{R}}
\newcommand{\sgn}{\textrm{sgn}}
\newcommand{\erf}{\textrm{erf}}

\newcommand{\bq}{\bbox{q}}
\newcommand{\hq}{\widehat{q}}

\newcommand{\bw}{\bbox{w}}
\newcommand{\bx}{\bbox{x}}
\newcommand{\hw}{\widehat{w}}
\newcommand{\hx}{\widehat{x}}
\newcommand{\bz}{\bbox{z}}

\begin{document}
\draft

\title{Dynamics of the  Batch Minority Game\\ with Inhomogeneous Decision Noise}
\author{A.C.C. Coolen$^\dag$, J.A.F. Heimel$^\dag$ and D. Sherrington$^\ddag$}
\address{\dag Department of Mathematics, King's College London, The Strand, London WC2R 2LS, UK\\
\ddag Department of Physics - Theoretical Physics, University of Oxford, 1 Keble Road, Oxford OX1 3NP, UK}
\date{\today}

\maketitle

\begin{abstract}
We study the dynamics of a version of the batch minority game,
with random external information and with different types of
inhomogeneous decision noise (additive and multiplicative), using
generating functional techniques \`{a} la De Dominicis. The
control parameters in this model are the ratio $\alpha=p/N$ of the
number $p$ of possible values for the external information over
the number $N$ of trading agents, and the statistical properties
of the agents' decision noise parameters. The presence of decision
noise is found to have the general effect of damping macroscopic
oscillations, which explains why in certain parameter regions it
can effectively reduce the market volatility, as observed in
earlier studies. In the limit $N\to\infty$ we (i) solve the first
few time steps of the dynamics (for any $\alpha$), (ii) calculate
the location $\alpha_c$ of the phase transition (signaling  the
onset of anomalous response), and (iii) solve the statics for
$\alpha>\alpha_c$. We find that $\alpha_c$ is not sensitive to
additive decision noise, but we arrive at non-trivial phase
diagrams in the case of multiplicative noise. Our theoretical
results find excellent confirmation in numerical simulations.
\end{abstract}


\section{Introduction}

One of the more recent application domains of equilibrium and
non-equilibrium statistical mechanics is the analysis of
simplified models describing large markets of competing traders
(or agents). One such model, which in spite of its apparent
simplicity was found to exhibit highly non-trivial behaviour and
has therefore attracted much attention, is the so-called Minority
Game (MG) \cite{ChalZhan97,Savit}, which is a variation on the
so-called El-Farol bar problem \cite{Arth94} which mimics in a
highly idealized fashion a market of speculators attempting to
profit by buying when most others wish to sell or selling when
others wish to buy, without individual knowledge of their fellows
but only of their collective consequences and external information
available to all. An extensive overview of the literature on the
MG and its many variations and extensions can be found in
\cite{Chalweb}. The striking feature of the MG, clearly observed
in numerical simulations, is the nontrivial dependence of the
market volatility (measuring global fluctuations) on the
dimensionality  of the information supplied to the agents (which
is defined as the relative number $\alpha$ of different values
which the information can take). For large $\alpha$ the volatility
approaches the  value corresponding to random trading, and the
system is ergodic. As $\alpha$ is reduced, also the volatility is
found to decrease beneath random, which is indicative of a more
efficient market, where agents have `learned' to improve the
effectiveness of their selection of trading strategies. A further
decrease of $\alpha$ will at some critical point $\alpha_c$ force
the system to undergo a phase transition to a highly non-ergodic
regime, where both a high-volatility state and a low-volatility
state can emerge, dependent on initial conditions (this was only
appreciated later).

In the original minority game, the information supplied to the
agents  consisted of the actual history of the market. However, it
was soon realized \cite{Cava99} that the dynamics of the MG
remains largely unaltered if, instead of the true history of the
market, random information is supplied to the agents; given
$\alpha$, the only relevant condition is that all agents must be
given the {\em same} information (whether sensible or otherwise).
This led to a considerable simplification of theoretical
approaches to the MG, since it reduced the process to a Markovian
one. A further generalization of the game was the introduction of
agents' decision noise \cite{CavaGarrGiarSher99}, which was shown
not only to improve worse than random behaviour but also, more
surprisingly, to be able to make it better than random. The study
\cite{CavaGarrGiarSher99} was followed by a number of papers
aiming to develop a solvable statistical mechanical theory, either
by using decision noise to `regularize' the stochastic equations
and replace these by deterministic ones (followed by an
equilibrium analysis of the ergodic regime, built on the
construction and exploitation of a Lyapunov function)
\cite{ChalMarsZecc00,MarsChalZecc00}, or by concentrating further
on analysis of the stochastic equations themselves
\cite{GarrMoroSher00}. Since the MG process does not obey detailed
balance, such studies (which also involved different
implementations of the decision noise), proved to be hard, and
their results partly controversial
\cite{ChalletPRL2000,CavagnaPRL2000} (especially with regard to
the questions of whether and when the stochastic MG equations can
be replaced by suitable deterministic ones).

More recently, in \cite{HeimelCoolen} the analysis of the MG was
approached from a different angle: all problems and debates
regarding microscopic determinism were simply avoided by
re-defining the MG dynamics directly in the form of discrete-time
deterministic equations, without decision noise (the so-called
`Batch Minority Game'). This allowed for an exact solution of the
model using generating functional techniques \`{a} la De Dominicis
\cite{DeDominicis}, which was found to be in excellent agreement
with numerical simulations, and which (due to it being dynamical
in nature) even applied to the non-ergodic regime. The present
study, which can be regarded as a natural follow-up on
\cite{HeimelCoolen}, achieves the following objectives. We
generalize the `Thermal Minority Game' such as to allow different
agents to have different levels of decision noise. This introduces
inhomogeneity into the agent population, which leads to
interesting new phenomena and phase diagrams. We generalize and
apply the (exact) formalism of \cite{HeimelCoolen} (which was
developed for the deterministic MG) to the case of having
inhomogeneous decision noise, within the context of the
discrete-time deterministic (`batch') equations. All our
theoretical results are shown to find excellent confirmation in
extensive numerical simulations.

\section{Model Definitions}

The minority game involves $N$ agents, labeled with Roman indices
$i,j,k,$ etc. At each round $\ell$ of the game, all agents act on
the basis of the same piece of external information $I(\ell)$.
 In the original
model \cite{ChalZhan97} the history of the actual market was used
as the information given to the agents. In view of the observation
in \cite{Cava99} that random information is equally efficacious we
here consider that at each round $\ell$ the agents are given the
information  $I(\ell)=I_{\mu(\ell)}$, where for each $\ell$ the
label $\mu(\ell)$ is chosen randomly and independently  from
$p=\alpha N$ possible values, i.e. $\mu(\ell)\in\{1,\ldots,\alpha
N\}$. To determine how to convert the external information into a
trading decision, each agent $i$ has at his/her disposal $S$
strategies $\bR_{ia}=(R_{ia}^1,\ldots,R_{ia}^{\alpha
N})\in\{-1,1\}^{\alpha N}$; $a\in\{1,\ldots,S\}$. Each component
$R_{ia}^\mu$ is selected randomly and independently from
$\{-1,1\}$ before the start of the game, with uniform
probabilities, and remains fixed throughout the game. The
strategies introduce quenched disorder into the model. Each
strategy $\bR_{ia}$ of every agent $i$ is given an initial
valuation or point-score $p_{ia}(0)$. In the deterministic version
of the game, given a choice $\mu(\ell)$ made for the information
presented at
 round $\ell$, every agent $i$ selects the strategy
which for trader $i$ has the highest  valuation at that point in
time, i.e. the strategy with label $\tilde{a}_i(\ell)=\mbox{arg
max}~ p_{ia}(\ell)$, and subsequently makes a binary bid
$b_i(\ell)=R^{\mu(\ell)}_{i\tilde{a}_i(\ell)}$. The (re-scaled)
total bid at stage $\ell$ is defined as $A(\ell)=N^{-1/2}\sum_i
b_i(\ell)$. Each agent subsequently updates the pay-off values of
each of his/her strategies $a$ on the basis of comparing the bid
which would have resulted from playing that strategy with the
actual outcome:
\begin{equation}
  p_{ia}(\ell\!+\!1)=p_{ia}(\ell) - R^{\mu(\ell)}_{ia} A(\ell).
\end{equation}
The minus sign in this expression ensures that strategies that
would have produced a minority decision are rewarded. Since the
qualitative behaviour of the market fluctuations was found to be
very much the same for all non-extensive numbers of strategies per
agent larger than one \cite{Cava99}, we restrict our discussion to
the $S=2$ model, where the equations can be simplified upon
introducing for each agent the instantaneous difference between
the two strategy valuations,
$q_i(\ell)=[p_{i1}(\ell)-p_{i2}(\ell)]/2$, as well as the average
strategy $\bomega_i=(\bR_{i1}+\bR_{i2})/2$ and the difference
between the strategies $\bxi_i=(\bR_{i1}-\bR_{i2})/2$. The
actually selected strategy in round $\ell$ can now be written
explicitly as a function of a binary variable $s_i(\ell)=\pm 1$,
which in the original model takes the value
$s_i(\ell)=\mbox{sgn}[q_i(\ell)]$, viz.
$\bR_{i\tilde{a}_i(\ell)}=\bomega_i+s_i(\ell)\bxi_i$, and the
evolution of the difference will now be given by:
\begin{equation}
\label{eq:online}
  q_i(\ell\!+\!1)=q_i(\ell)-\xi_i^{\mu(\ell)}[
    \Omega^{\mu(\ell)} \! + \!
    \frac{1}{\sqrt{N}}\sum_j \xi_j^{\mu(\ell)} s_j(\ell)
  ],
\end{equation}
with $\bOmega=N^{-1/2}\sum_j \bomega_j\in \R^{\alpha N}$.

In the so-called thermal minority game \cite{CavaGarrGiarSher99},
the deterministic decision rule $s_j(\ell)=\sgn[q_j(\ell)]$ was
replaced by a stochastic recipe; another recipe was employed in
\cite{GarrMoroSher00}. Here we generalise this idea further, by
allowing different traders to have different levels of
stochasticity in their decision making. We will consider decision
noise of the general form
\begin{equation}
s_j(\ell)=\sigma[q_j(\ell),z_j(\ell)|T_j],
 \label{eq:generalnoise}
\end{equation}
in which the $z_{j}(\ell)$ are independent and zero-average random
numbers, described by some symmetric distribution $P(z)$ which is
normalised according to
 $\int\!dz~P(z)=\int\!dz~P(z)z^2=1$. The function $\sigma[q,z|T]\in\{-1,1\}$ is
 chosen to interpolate smoothly via a single control parameter $T$
 between
 $\sigma[q,z|0]=\sgn[q]$ for $T=0$ and
$\sigma[q,z|\infty]=\pm 1$ (randomly, with equal probabilities)
for $T=\infty$, so that $T$ provides a measure
 of the degree of stochasticity in the traders' decision making
(with random choice in the case $q=0$). Typical examples are
 additive and multiplicative noise definitions such as
\begin{eqnarray}
{\rm additive:~~~~~~~~~~} && \sigma[q,z|T]=\sgn[q+Tz]
\label{eq:additive}
\\
{\rm multiplicative:~~~} && \sigma[q,z|T]=\sgn[q]~\sgn[1+Tz].
\label{eq:multiplicative}
\end{eqnarray}
 In the first case (\ref{eq:additive}) the noise has the potential to be overruled by the
 so-called `frozen' agents \cite{Savit}, who have $q_i(t)\sim \tilde{q}_i t$
 for $t\to\infty$ \cite{ChalletPRL2000,CavagnaPRL2000,HeimelCoolen}.
In the second case the decision noise will even retain its effect
for `frozen' agents (if they exist).
 The
above definitions represent situations where for $T_i>0$ a trader
$i$ need not always use his/her `best' strategy; for $T_i\to 0$ we
revert back to the deterministic model. The impact of the
multiplicative noise (\ref{eq:multiplicative}) can be
characterised by the monotonic function
\begin{equation}
\lambda(T)=\int\!dz~P(z)~\sgn[1+Tz],
 \label{eq:lambdaT}
\end{equation}
with $\lambda(0)=1$ and $\lambda(\infty)=0$. For example, for  a
Gaussian $P(z)$ one has $\lambda(T)=\erf[1/\sqrt{2}T]$.

Finally in the formulation of the model, we replace the above
`on-line' version of the microscopic dynamics (\ref{eq:online}),
following \cite{HeimelCoolen}, by a so-called `batch' version,
where, rather than modifying the $\{q_i\}$ after every observation
of an individual piece of external information, they are modified
according to the {\em average} effect of the possible choices for
the external information:
\begin{equation}
q_i(\ell\!+\!1)=q_i(\ell)-\frac{1}{p}\sum_{\mu=1}^p \xi_i^{\mu}[
    \Omega^{\mu} \! + \!
    \frac{1}{\sqrt{N}}\sum_j \xi_j^{\mu} s_j(\ell)],
\end{equation}
giving
\begin{equation}
q_i(t+1)=q_i(t)- h_i - \sum_j J_{ij} ~\sigma[q_j(t),z_j(t)|T_j],
  \label{eq:batch}
\end{equation}
where $J_{ij}=2N^{-1}\bxi_i\cdot\bxi_j$ and
$h_i=2N^{-\frac{1}{2}}\bxi_i\cdot\bOmega$. The specific choice of
time scaling in (\ref{eq:batch}) has been made for later
convenience. The batch dynamics (\ref{eq:batch}) has the advantage
of being sufficiently simple and transparent to allow for a
straightforward exact dynamical solution of the model, using
generating functional techniques \cite{HeimelCoolen}. The process
(\ref{eq:batch}) is not exactly equivalent to (\ref{eq:online}),
not even for $N\to\infty$ (see \cite{Coolen} for the generating
functional analysis of the on-line dynamics and its relation to
the batch alternative), but it does present qualitatively similar
features.

There are many ways to introduce a stochastic element into the
traders' decision making. For instance, the two versions of the
minority game studied in \cite{CavaGarrGiarSher99,GarrMoroSher00}
correspond to the forms (\ref{eq:additive}) and
(\ref{eq:multiplicative}) with $P(z)=\frac{1}{2}K[1-\tanh^2(Kz)]$
and $T_i=T$ for all $i$ (giving strategy selection probabilities
of the form ${\rm Prob}(\sigma=\pm 1)\sim e^{\pm\beta q}$ and
${\rm Prob}(\sigma=\pm 1)\sim e^{\pm\beta \sgn[q]}$,
respectively).

The magnitude of the market fluctuations, or volatility, is given
by
\begin{equation}
\sigma^2=\bra\frac{1}{p}\sum_\mu(A^\mu)^2\ket_{\bz}-\bra\frac{1}{p}\sum_\mu
A^\mu\ket_z^2, \label{eq:define_volatility}
\end{equation}
 where
$A^\mu=N^{-\frac{1}{2}}\sum_i[\omega_i^{\mu}+s_i\xi_i^{\mu}]$ and
where  $\bra\ldots\ket_{\bz}$ denotes an average over the random
numbers $\{z_i\}$. One easily derives
\begin{eqnarray}
\bra\frac{1}{p}\sum_\mu A^\mu\ket_{\bz}&=&\frac{1}{\alpha
N\sqrt{N}}\sum_i \bra s_i\ket_{\bz}
\sum_\mu\xi_i^\mu+\order(N^{-\frac{1}{2}}), \label{eq:average}
\\
\bra \frac{1}{p}\sum_\mu
(A^\mu)^2\ket_{\bz}&=&\frac{1}{2}+\frac{1}{\alpha N}[\sum_i h_i
\bra s_i\ket_{\bz}
   + \frac{1}{2}\sum_{ij} J_{ij} \bra s_i s_j\ket_{\bz} ]
\nonumber\\ && \hspace*{15mm}
 +\order(N^{-\frac{1}{2}}).
 \label{eq:volatility}
 \end{eqnarray}
 Purely random trading corresponds to $\bra\frac{1}{p}\sum_\mu A\ket_{\bz}=0$ and
 $\sigma^2=1$.
Following \cite{HeimelCoolen} we also define the volatility matrix
$\Xi_{tt^\prime}$:
\begin{equation}
\Xi_{tt^\prime}=\bra \frac{1}{p}\sum_\mu [A^\mu_t-\bra
\frac{1}{p}\sum_\nu A^\nu_t\ket_{\bz}][A^\mu_{t^\prime}-\bra
\frac{1}{p}\sum_\nu A^\nu_{t^\prime}\ket_{\bz}]\ket_{\bz},
 \label{eq:volatility_matrix}
\end{equation}
which measures the temporal correlations of the market fluctuations.
Note that $\sigma^2_t=\Xi_{tt}$.
In the case where
 the average bid $\langle A \rangle$ is zero (as in the present
 model), the volatility measures the efficiency of the market.

\section{Generating Functional Analysis}

The canonical tool to deal with the dynamics of the present
problem is generating functional analysis \`{a} la De Dominicis
\cite{DeDominicis}, which allows one to carry out the disorder
average (here: the average over all strategies) and take the
$N\to\infty$ limit exactly. The final result of the analysis is a
set of closed equations, which can be interpreted as describing
the dynamics of an effective `single agent'
\cite{DeDominicis,SompZipp82}. Due to the disorder in the process,
this single agent will acquire an effective `memory', i.e. he/she
will evolve according to a non-trivial non-Markovian stochastic
process. Here we will follow closely the steps taken in
\cite{HeimelCoolen}, and we refer to the latter paper for full
details of the calculation. In contrast to the situation in
\cite{HeimelCoolen}), for the present noisy version of the game
one finds a microscopic transition probability density operator
$W(\bq|\bq^\prime)$:
\begin{eqnarray}
  W(\bq|\bq')
  &=&
  \int\!\!  \frac{d\hat{\bq}}{(2\pi)^N}~
\bra  e^{\sum_i i\hq_i (
     q_i - q_i' +
   h_i + \sum_j J_{ij} s^\prime_j)
  }
  \ket_{\bz},
\end{eqnarray}
with the short-hand $s_j^\prime=\sigma[q_j^\prime,z_j|T_j]$.
The moment generating functional for a stochastic process of the
present type is defined as
\begin{eqnarray}
  Z[\bpsi]
  &=&
  \langle~e^{i\sum_t \sum_i  \psi_i(t) q_i(t)}~\rangle
  \nonumber \\
  &=&
  \int\prod_t\left[ d\bq(t)~ W(\bq(t+1)|\bq(t))\right]~p_0(\bq(0))
  \nonumber \\
  &&
   \times ~e^{i\sum_t \sum_i  \psi_i(t) q_i(t)}.
\end{eqnarray}
Derivation of the generating functional with respect to the
conjugate variables $\bpsi$ generates all moments of $\bq$ at
arbitrary times. Upon introducing the two short-hands:
\begin{equation}
  w^\mu_t=\frac{\sqrt{2}}{\sqrt{N}}\sum_i \hq_i(t) \xi^\mu_i, \qquad
  x^\mu_t=\frac{\sqrt{2}}{\sqrt{N}}\sum_i s_i(t)\xi^\mu_i,
\end{equation}
as well as
$D\bq=\prod_{it}[dq_i(t)/\sqrt{2\pi}]$,
$D\bw=\prod_{\mu t}[dw^\mu_t/\sqrt{2\pi}]$ and
$D\bx=\prod_{\mu t}[dx^\mu_t/\sqrt{2\pi}]$
(with similar definitions for $D\hat{\bq}$, $D\hat{\bw}$ and $D\hat{\bx}$, respectively),
the generating functional takes the following form:
\begin{eqnarray}
  Z[\bpsi]
  &=&
  \int\! D\bw D\hat{\bw} D\bx D\hat{\bx}~
  e^{i\sum_{t \mu}[\hw^\mu_t w^\mu_t +
        \hx^\mu_t x^\mu_t +  \sqrt{2} w^\mu_t(\Omega^\mu +
        x^\mu_t)]} \nonumber
  \\&&\hspace*{-5mm}
  \times \int\! D\bq D\hat{\bq}~p_0(\bq(0))~
  \bra e^{\frac{-i\sqrt{2}}{\sqrt{N}}\sum_{\mu i}\xi_i^\mu\sum_t [
      \hw^\mu_t \hq_i(t) + \hx^\mu_t s_i(t) ]}\ket_{\bz}
       \nonumber
  \\&&\hspace*{-5mm}
  \times ~e^{i\sum_{t i}\left[\hq_i(t) (
     q_i(t+1) - q_i(t)-\theta_i(t))+\psi_i(t) q_i(t)\right]},
 \label{eq:Zbeforeaverage}
\end{eqnarray}
where, as in \cite{HeimelCoolen}, we introduced external `forces'
$\theta_i(t)$ to generate response functions.

To describe typical behaviour, and in view of the self-averaging
character of the large $N$ limit, at this stage we average over
over the explicit choices of the quenched random parameters
$\{\bR\}$. These averages are not affected in any way by the
introduction of the noise variables $\{z_i\}$ or the independent
temperatures $T_i$, and the further procedure of
\cite{HeimelCoolen} still applies here, generating the dynamical
order parameters $C_{t t^\prime}=N^{-1}\sum_i s_i(t)
s_i(t^\prime)$, $K_{t t^\prime}=N^{-1}\sum_i s_i(t)
\hq_i(t^\prime)$, and $L_{t t^\prime}=N^{-1}\sum_i \hq_i(t)
\hq_i(t^\prime)$ and their conjugates. For times which do not
scale with $N$ and for simple initial conditions of the form
$p_0(\bq)=\prod_i p_{0}(q_i)$ one thus finds:
\begin{equation}
  \overline{Z[\bpsi]}=
  \int\![DC D\hat{C}][DK D\hat{K}][ DL D\hat{L}]~
    e^{N\left[\Psi+\Phi+\Omega \right]+\order(N^0)}.
    \label{eq:Zafteraverage}
\end{equation}
The $\order(N^0)$ term in the exponent is
independent of the fields $\{\psi_i(t)\}$ and
$\{\theta_i(t)\}$.
The three relevant exponents in (\ref{eq:Zafteraverage}) are given by
the following expressions:
\begin{eqnarray}
\Psi&=&i\sum_{tt^\prime}[ \hC_{tt^\prime} C_{tt^\prime} +\hK_{t t^\prime} K_{tt^\prime}
+ \hL_{tt^\prime} L_{tt^\prime}],
\label{eq:Psi}
\\
\Phi &=& \alpha  \ln\left[ \int\! Dw D\hat{w} Dx D\hat{x}~
  e^{i\sum_{t}[\hw_t w_t + \hx_t x_t +  w_t x_t]}
  \right.
\nonumber\\
&&
\hspace*{-3mm}
\times
\left.
e^{-\frac{1}{2}\sum_{t t^\prime} \left[
 w_t w_{t^\prime}
 +\hw_t L_{t t^\prime} \hw_{t^\prime}
 +2\hx_{t} K_{t t^\prime} \hw_{t^\prime}
 +\hx_t C_{t t^\prime}\hx_{t^\prime}
\right]}\right]
\label{eq:Phi}
\\
\Omega &=& \frac{1}{N}\!\sum_i\ln \left[ \int\! Dq
D\hat{q}~p_0(q(0)) \nonumber\right.\\ && \left.\hspace*{-3mm}
\times~ e^{i\sum_{t}\left[\hq(t)[
     q(t+1) - q(t)-\theta_i(t)]+\psi_i(t) q(t)\right]-i\sum_{t t^\prime}\hq(t)\hat{L}_{t t^\prime}  \hq(t^\prime)}
\right.
\nonumber \\
&&
\left.\hspace*{-3mm}
\times~\bra
e^{-i\sum_{t t^\prime}[s_i(t) \hat{C}_{t t^\prime}s_i(t^\prime)
+s_i(t)\hat{K}_{t t^\prime}  \hq(t^\prime)]}\ket_{\bz}
     \right].
\label{eq:Omega}
\end{eqnarray}
Here $s_i(t)=\sigma[q(t),z_t|T_i]$ and the average $\bra
\ldots\ket_{\bz}$ has now been reduced to a single site (but
many-time) one: $\bra g[z_1,z_2,\ldots]\ket_{\bz} =\int\!\prod_t
[dz_t P(z_t)]~g[z_1,z_2,\ldots]$. Following \cite{HeimelCoolen} we
have also introduced the short-hands
$Dq=\prod_{t}[dq(t)/\sqrt{2\pi}]$,
$Dw=\prod_{t}[dw_t/\sqrt{2\pi}]$, $Dx=\prod_{t}[dx_t/\sqrt{2\pi}]$
(with similar definitions for $D\hat{q}$, $D\hat{w}$ and
$D\hat{x}$). Note that all the quantities appearing in
(\ref{eq:Zafteraverage}) are macroscopic; all the microscopic
variables have been integrated out.

\section{The Saddle-Point Equations}

We can now evaluate (\ref{eq:Zafteraverage}) by saddle-point
integration, in the limit $N\to\infty$. We define $G_{t
t^\prime}=-iK_{t t^\prime}$. Taking derivatives with respect to
the generating fields
 and using the normalisation $\overline{Z[{\bf
0}]}=1$ then gives (at the physical saddle-point) the usual
identifications
\begin{eqnarray}
C_{t t^\prime}&=&\lim_{N\to\infty}\frac{1}{N}\sum_i\overline{\bra
s_i(t) s_i(t^\prime)\ket},
 \label{eq:meaningof_C}
\\
G_{t
t^\prime}&=&\lim_{N\to\infty}\frac{1}{N}\sum_i\frac{\partial}{\partial
\theta_i(t^\prime)}\overline{\bra s_i(t)\ket},
\label{eq:meaningof_K}
\end{eqnarray}
and also
\begin{eqnarray}
L_{t t^\prime}&=&-\lim_{N\to\infty}\frac{1}{N}\sum_i
\frac{\partial^2}{\partial\theta_i(t)\partial
\theta_i(t^\prime)}1= 0.
 \label{eq:meaningof_L}
\end{eqnarray}
Putting $\psi_i(t)=0$ (they are no longer needed) and $\theta_i(t)=\theta(t)$
then simplifies (\ref{eq:Omega}) to
\begin{eqnarray}
\Omega &=&\int_0^\infty\!\!dT~W(T)\ln \left[ \int\! Dq
D\hat{q}~p_0(q(0))~ \nonumber\right.
\\
&&
\left.
\times~
e^{i\sum_{t}\hq(t)[
     q(t+1) - q(t)-\theta(t)]
     -i\sum_{t t^\prime}\hq(t)\hat{L}_{t t^\prime}  \hq(t^\prime)}
\nonumber\right.\\
&&
\left.
\times~
\bra e^{-i\sum_{t t^\prime}[s(t) \hat{C}_{t
t^\prime}s(t^\prime)
+s(t)\hat{K}_{t t^\prime}  \hq(t^\prime)]}
\ket_{\bz}
     \right],
\label{eq:Omega_new}
\end{eqnarray}
in which now $s(t)=\sigma[q(t),z_t|T]$, and where $W(T)$ denotes the distribution of local noise
strengths:
\begin{equation}
W(T)=\lim_{N\to\infty} \frac{1}{N}\sum_i\delta[T-T_i].
\label{eq:temp_dist}
\end{equation}
Extremisation of the extensive exponent $N[\Psi+\Phi+\Omega]$ of
(\ref{eq:Zafteraverage}) with respect to
$\{C,\hat{C},K,\hat{K},L,\hat{L}\}$ gives the  saddle-point
equations
\begin{eqnarray}
C_{t t^\prime}=\langle s(t) s(t^\prime) \rangle_\star ~~~~~~~~
G_{t t^\prime}=\frac{\partial \bra
s(t)\ket_\star}{\partial\theta({t^\prime})}, ~~~~~~~~
\label{eq:CandG}
\\
\hC_{tt^\prime}=\frac{i\partial \Phi}{\partial C_{t t^\prime}}
~~~~~~ \hK_{tt^\prime}=\frac{i\partial \Phi}{\partial K_{t
t^\prime}} ~~~~~~ \hL_{tt^\prime}=\frac{i\partial \Phi}{\partial
L_{t t^\prime}}.
 \label{eq:Conjugates}
\end{eqnarray}
The effective single-trader
averages $\bra \ldots\ket_\star$, generated by taking
derivatives of (\ref{eq:Omega}), are defined as
\begin{equation}
\bra f[\{q,s\}]\ket_\star= \int_0^\infty\!\!dT~W(T)\left\{
\frac{\int\! Dq ~\bra M[\{q,s\}]f[\{q,s\}]\ket_{\bz}}{\int\! Dq
~\bra M[\{q,s\}]\ket_{\bz}} \right\},
 \label{eq:effective_measure}
\end{equation}
\vspace*{-3mm}
\begin{eqnarray}
M[\{q,s\}]&=& p_0(q(0))~e^{-i\sum_{t t^\prime} s(t) \hat{C}_{t
t^\prime}s(t^\prime)} \nonumber \\ &&\times \int\!
D\hat{q}~e^{-i\sum_{t t^\prime}\hq(t)\hat{L}_{t
t^\prime}\hq(t^\prime)} \nonumber \\ &&\times~
e^{i\sum_{t}\hq(t)[q(t+1) -
q(t)-\theta(t)-\sum_{t^\prime}\hat{K}^T_{t t^\prime}s(t^\prime)]},
\label{eq:singletrader_measure}
\end{eqnarray}
Upon elimination of the trio $\{\hat{C},\hat{K},\hat{L}\}$ via
(\ref{eq:Conjugates}) we obtain exact closed equations for the
disorder-averaged correlation- and response functions in the
$N\to\infty$ limit: equations (\ref{eq:CandG}), with the effective
single trader measure (\ref{eq:singletrader_measure}). One
recovers the theory of \cite{HeimelCoolen} upon putting
$W(T)=\delta(T)$.

Since the introduction of decision noise into the dynamics has
only affected the term $\Omega$ (\ref{eq:Omega_new}), compared to
the analysis in \cite{HeimelCoolen}, the simplifications of the
term $\Phi$ (reflecting the statistical properties of the trading
strategies) derived in \cite{HeimelCoolen} apply unaltered, so
that at the physical saddle-point we again find
\begin{eqnarray}
\hL&=&-\frac{1}{2}i\alpha
    (\openone+G)^{-1}
    D(\openone+G^T)^{-1},
\label{eq:hatL}
\\
\hK^T&=&-\alpha(\openone+G)^{-1},
 \label{eq:hatK}
\\
\hC&=&0,
\label{eq:hatC}
\end{eqnarray}
where $A^T$ denotes the transpose of the matrix $A$, and
the entries of the matrix $D$ are given by $D_{tt^\prime}= 1+C_{tt^\prime}$.
We now find
our effective single trader measure $M[\{q,s\}]$ of
(\ref{eq:singletrader_measure}) reducing further to
\begin{eqnarray}
M[\{q,s\}]&=& p_0(q(0)) \nonumber
\\
&& \hspace*{-10mm}
 \times
 \int\!\!
D\hat{q}~e^{-\frac{1}{2}\alpha\sum_{t
t^\prime}\hq(t)[(\openone+G)^{-1} D(\openone+G^T)^{-1}]_{t
t^\prime} \hq(t^\prime)} \nonumber \\ && \hspace*{-10mm} \times~
e^{i\sum_{t}\hq(t)[q(t+1) -
q(t)-\theta(t)+\alpha\sum_{t^\prime}(\openone+G)^{-1}_{t t^\prime}
s(t^\prime)]}.
 \label{eq:singleagent_statistics}
\end{eqnarray}
For a given value of $T$,
this describes a stochastic single-agent process of the form
\begin{eqnarray}
  q(t\!+\!1)& =& q(t) - \alpha \sum_{t^\prime\leq t}
  (\openone+ G)^{-1}_{t t^\prime} \sigma[q(t^\prime),z_{t^\prime}|T]
\nonumber \\ &&  +~\theta(t) +\sqrt{\alpha}~\eta(t).
\label{eq:singleagent}
\end{eqnarray}
Causality ensures that $(\openone+ G)^{-1}_{tt^\prime}=0$ for $t^\prime >t$. The variable
$z_t$ represents the original single-trader decision noise,
with $\bra z_t\ket=0$ and $\bra z_t z_{t^\prime}\ket=\delta_{tt^\prime}$, and
 $\eta(t)$ is a disorder-generated Gaussian noise with zero mean and with
  temporal correlations given by
  $\bra\eta(t)\eta(t^\prime)\ket=\Sigma_{tt^\prime}$:
\begin{equation}
\Sigma = (\openone+G)^{-1} D(\openone+G^T)^{-1}.
\label{eq:noise_covariance}
\end{equation}
The correlation- and response functions (\ref{eq:meaningof_C},\ref{eq:meaningof_K}) are the dynamic
order parameters of the problem, and must be solved
self-consistently from the closed equations
\begin{eqnarray}
C_{t t^\prime}&=&\bra
~\sigma[q(t),z_t|T]~\sigma[q(t^\prime),z_{t^\prime}|T]~\ket_\star,
\label{eq:finalC}
\\[1mm]
G_{t t^\prime}&=&\frac{\partial }{\partial\theta({t^\prime})}
\bra~ \sigma[q(t),z_t|T]~\ket_\star,
\label{eq:finalG}
\end{eqnarray}
which, following (\ref{eq:effective_measure}), now also involve
averaging over the distribution of the noise strengths $T$.
Note that $M[\{q,s\}]$ as given by
(\ref{eq:singleagent_statistics})
is normalised, i.e. $\int\! Dq~M[\{q,s\}]=1$, so
 the associated averages reduce to
\begin{equation}
\bra f[\{q,s\}]\ket_\star=\int_0^\infty\!\!dT~W(T)\int\! Dq ~\bra
M[\{q,s\}]f[\{q,s\}]\ket_{\bz}.
 \label{eq:new_effective_measure}
\end{equation}
The calculation in \cite{HeimelCoolen} of the disorder-averaged
average bid and the volatility
matrix (including the single-time volatility
$\sigma_t^2=\Xi_{tt}$) still hold, and hence
\begin{equation}
\lim_{N\to\infty}\overline{\bra A\ket_t}=0,~~~~~~~~
\lim_{N\to\infty}\overline{\Xi}_{tt^\prime}=\frac{1}{2}\Sigma_{t
t^\prime}.
\label{eq:results_of_appendix}
\end{equation}

\section{The First Time Steps}

For the first few time steps one can calculate quite easily
the order parameters (correlation- and response functions) and the volatility,
from (\ref{eq:singleagent_statistics}), using  the simplifications which follow from
causality such as
\begin{equation}
[G^{n}]_{tt^\prime}=0~~~{\rm for}~~~t^\prime>t-n.
\label{eq:causality}
\end{equation}
At $t=0$ this immediately allows us to conclude that
$\Sigma_{00}=D_{00}=2$.
We now obtain from (\ref{eq:singleagent_statistics}) the joint statistics at times $t=1$, given a value for
$T$:
\begin{eqnarray}
p(q(1)|q(0))&=& \nonumber\\ &&\hspace*{-20mm}
\int\!dz_0~P(z_0)\frac{e^{-\left[q(1)-q(0)-\theta(0)+\alpha~
\sigma[q(0),z_0|T]\right]^2/4\alpha}}{2\sqrt{\alpha \pi}}.
\label{eq:joint_dist_t=01}
\end{eqnarray}
Equation (\ref{eq:joint_dist_t=01}) allows us to calculate $C_{10}$
and $G_{10}$, although the presence of the decision noise
induces expressions which are significantly more difficult to work
out explicitly than those of the noise-free case in
\cite{HeimelCoolen}, and which will depend on the choice
made for $\sigma[q,z|T]$:
\begin{eqnarray}
C_{10}&=&\int_0^\infty\!\!dT~W(T)\int\!dz_0dz_1 P(z_0)P(z_1)
\int\!dq(0)p_0(q(0)) \nonumber\\ &&
\times~\int\!\frac{dq(1)}{2\sqrt{\alpha \pi}}~
e^{-\left[q(1)-q(0)-\theta(0)+\alpha~
\sigma[q(0),z_0|T]\right]^2/4\alpha} \nonumber\\ && \times~
\sigma[q(0),z_0|T]~\sigma[q(1),z_1|T],
 \label{eq:C10}
\\
G_{10}&=&\int_0^\infty\!\!dT~W(T)\int\!dz_0dz_1 P(z_0)P(z_1)
\int\!dq(0)p_0(q(0)) \nonumber\\ &&
\times~\int\!\frac{dq(1)}{2\sqrt{\alpha \pi}}~
e^{-\left[q(1)-q(0)-\theta(0)+\alpha~
\sigma[q(0),z_0|T]\right]^2/4\alpha} \nonumber\\ && \times~
\frac{\partial}{\partial q(1)}\sigma[q(1),z_1|T].
 \label{eq:G10}
\end{eqnarray}
We can now move to the next time step, again using (\ref{eq:causality}), where we need the noise
covariances $\Sigma_{11}$ and $\Sigma_{10}$:
\begin{eqnarray}
\Sigma_{10}&=& 1+C_{10}-2G_{10},
 \label{eq:S10}
\\[1mm]
\Sigma_{11} &=& 2-2G_{10}[1+ C_{01}] +2[G_{10}]^2.
 \label{eq:S11}
\end{eqnarray}
This procedure can in principle be repeated for an arbitrary number of time steps.
\vsp

We now specialise to the case where the game  is initialised
in a \emph{tabula rasa} manner,
i.e. $p(q(0))=\delta[q_0]$, and where we have no perturbation fields, i.e. $\theta(t)=0$.
Now, upon also using the symmetry of $P(z)$, we can reduce the above results to
\begin{eqnarray}
C_{10}&=&\int_0^\infty\!\!dT~W(T)\int\!dz~
P(z)\int\!\frac{dq}{4\sqrt{\alpha \pi}}~
e^{-\left[q+\alpha\right]^2/4\alpha} \nonumber\\ && \times~
\left\{\sigma[q,z|T]-\sigma[-q,-z|T]\right\}.
\\
G_{10}&=&\int_0^\infty\!\!dT~W(T)\int\!dz~ P(z)
\int\!\frac{dq}{4\sqrt{\alpha \pi}}~
e^{-\left[q+\alpha\right]^2/4\alpha} \nonumber\\ &&
\times~\frac{\partial}{\partial q}
\left\{\sigma[q,z|T]-\sigma[-q,-z|T]\right\}.
\end{eqnarray}
Inspection of these expressions for large and small $\alpha$, and
for the specific choices
(\ref{eq:additive},\ref{eq:multiplicative})
reveals the following. For $\alpha\to \infty$ one finds
\begin{eqnarray}
&&\lim_{\alpha\to\infty} G_{10}=0,~~~~~~
 \lim_{\alpha\to\infty}\Sigma_{11}=2,
\end{eqnarray}
for both noise types. The order parameters $C_{10}$ and $\Sigma$,
in contrast, are sensitive to the type of noise chosen. For
additive noise of the form (\ref{eq:additive}) one has
\begin{eqnarray}
&&\lim_{\alpha\to \infty} C_{10}=-1,~~~~~~
\lim_{\alpha\to\infty}\Sigma_{10}=0,
\end{eqnarray}
whereas for multiplicative noise (\ref{eq:multiplicative}) one has
\begin{eqnarray}
\lim_{\alpha\to \infty}
C_{10}&=&-\int_0^\infty\!\!dT~W(T)\lambda(T),
\\
\lim_{\alpha\to\infty}\Sigma_{10}&=&1-\int_0^\infty\!\!dT~W(T)\lambda(T).
\end{eqnarray}
In both cases the negativity of $C_{10}$ shows that the {\em
tabula-rasa} initialised system immediately enters an oscillation,
with the $q_i(1)$ on average having opposite sign to the
corresponding $q_i(0)$. Initially, additive noise is found not to
play a role, and the effective disorder-generated noise components
$\eta(t)$ decorrelate, compared with the deterministic case of
\cite{HeimelCoolen}. Multiplicative noise, on the other hand, is
seen to retain an impact, even for short times and large $\alpha$,
and to cause a reduction of the oscillation amplitude.

Now we turn to small
$\alpha$, where we make the choice $P(z)=(2\pi)^{-\frac{1}{2}}e^{-z^2/2}$
in order to work out integrals explicitly. For additive noise (\ref{eq:additive}) we find
\begin{eqnarray}
C_{10} &=&-\frac{\alpha\sqrt{2}}{\sqrt{\pi}}
\int_0^\infty\!\!dT~W(T)~T^{-1}+\order(\alpha^{\frac{3}{2}}),
\\
G_{10}&=&\frac{\sqrt{2}}{\sqrt{\pi}}\int_0^\infty\!\!dT~W(T)~T^{-1}
+\order(\alpha^{\frac{1}{2}})
\end{eqnarray}
(provided the above integrals over $T$ exist; if they do not, we
revert back to the leading orders of the $T=0$ case
\cite{HeimelCoolen}, i.e. $C_{10}=\order(\sqrt{\alpha})$ and
$G_{10}=\order(1/\sqrt{\alpha})$).
 Combination with the expressions
(\ref{eq:S10},\ref{eq:S11}) shows that in leading order
\begin{equation}
\eta(1)=(\frac{1}{2}-G_{10})\eta(0)+w+\ldots
\end{equation}
in which $w$ is a zero-average Gaussian variable, independent of
$\eta(0)$,
with variance $\bra w^2\ket=3/2$.
Hence we find from the effective single spin equation
(\ref{eq:singleagent}):
\begin{eqnarray}
  q(1)& =& \sqrt{\alpha}~\eta(0)+\order(\alpha)
\\
  q(2)& =& \sqrt{\alpha}\left[
  (\frac{3}{2}-G_{10})\eta(0)+w\right]+\order(\alpha).
\end{eqnarray}
We observe, as in \cite{HeimelCoolen}, that for small $\alpha$ and
additive decision noise the first two time steps are driven
predominantly by the disorder-generated noise component in
(\ref{eq:singleagent}). However, whether this noise component
starts oscillating in sign is, in the case of decision noise,
crucially dependent on the distribution of temperatures; only when
$\int\!\!dT~W(T) ~T^{-1}$ is sufficiently large should we expect
the system to enter the high volatility state. For multiplicative
noise, on the other hand, we arrive for small $\alpha$ at the
leading orders
\begin{eqnarray}
C_{10}&=&-\frac{\sqrt{\alpha}}{\sqrt{\pi}}\int_0^\infty\!\!dT~W(T)\lambda(T)
+\order(\alpha^{\frac{3}{2}}),
\\
G_{10}&=&\frac{1}{\sqrt{\alpha
\pi}}\int_0^\infty\!\!dT~W(T)\lambda(T)+\order(\sqrt{\alpha})
+\ldots
\end{eqnarray}
Here the oscillation is much stronger (provided we do not scale the temperatures with $\alpha$).
Combination with the expressions (\ref{eq:S10},\ref{eq:S11}) shows
that in leading order the disorder-generated noise not only drives the oscillation,
but is also being amplified by a factor of order $\alpha^{-1/2}$:
\begin{equation}
\eta(1)=-G_{10}\eta(0)+\order(\alpha^0)
\end{equation}
The effective single trader equation subsequently gives:
\begin{eqnarray}
q(1)& =& \sqrt{\alpha}~\eta(0)+\order(\alpha),
  \\
q(2)& =&
  -\frac{\eta(0)}{\sqrt{\pi}}\int_0^\infty\!\!dT~W(T)\lambda(T)
  +\order(\sqrt{\alpha}).
\end{eqnarray}
Thus, for small $\alpha$  and tabula rasa
initialisation\footnote{Note that the small $\alpha$ expansions in
this section are made for fixed $W(T)$; the observed behaviour is
likely to be different when $W(T)$ is allowed to scale with
$\alpha$.} additive decision noise has the most drastic effect on
the dynamics, changing the leading order of the relevant
observables by a factor $\sqrt{\alpha}$ (in contrast to
multiplicative noise).

\section{Stationary State for $\alpha>\alpha_c(W(T))$}

If the game has reached a time-translation invariant stationary
state without long-term memory, then
$G_{tt^\prime}=G(t-t^\prime)$, $C_{tt^\prime}=C(t-t^\prime)$ and
$\Sigma_{t t^\prime}=\Sigma(t-t^\prime)$. In this section we
assume that the stationary state is one without anomalous
response, i.e. $\lim_{\tau\rightarrow\infty}\sum_{t\leq \tau}
G(t)=k$ exists. The lower limit of such behaviour in $\alpha$
defines $\alpha_c(W(T))$.

In a stationary state one generally finds agents who change
strategy frequently, but also agents who are consistently in the
minority group. For the latter `frozen' agents, the  values of
$q_i$ will grow linearly in time. We follow \cite{HeimelCoolen}
and separate the two groups by introducing
$\tilde{q}_i(t)=q_i(t)/t$;  frozen agents will be those for whom
$\lim_{t\to\infty}\tilde{q}_i(t)\neq 0$, and the quantity
$\phi=\lim_{\epsilon\to 0}\lim_{t\to\infty}\bra
\theta[|\tilde{q}(t)|-\epsilon]\ket_\star$ gives the fraction of
`frozen' agents in the original $N$-agent system, for $N\to
\infty$. Transformation of the process (\ref{eq:singleagent})
gives, for a given $T$:
\begin{eqnarray}
\widetilde{q}_T(t)&=&
  \frac{1}{t}\widetilde{q}_T(1)
 + \frac{\sqrt{\alpha}}{t}\sum_{t^\prime <t}\eta(t^\prime)
 \nonumber\\
&& -\frac{\alpha}{t}\sum_{t^\prime <t}
   \sum_{t^\pprime\leq t^\prime}(\openone+G)^{-1}_{t^\prime
   t^\pprime}~\sigma[\tilde{q}_T(t^\pprime),z_{t^\pprime}|T].
\label{eq:rescaled_agent}
\end{eqnarray}
We now define $\tilde{q}_T=\lim_{t\to\infty}\tilde{q}_T(t)$
(assuming this limit exists) and take the limit $t\to\infty$ in (\ref{eq:rescaled_agent}),
giving
\begin{equation}
  \widetilde{q}_T=-\frac{\alpha}{1+k}m_T+\sqrt{\alpha}~\eta,
\label{eq:stationarity}
\end{equation}
with the time averages $m_T=\lim_{\tau\to\infty}\frac{1}{\tau}
\sum_{t\leq
  \tau}\sigma[q_t,z_t|T]$\\ and
  $\eta=\lim_{\tau\to\infty}\frac{1}{\tau}\sum_{t\leq
  \tau}\eta(t)$.
The variance of $\eta$
follows from
(\ref{eq:noise_covariance}):
\begin{eqnarray}
\bra \eta^2 \ket &=&
(1+k)^{-2}[1+\lim_{\tau,\tau^\prime\to\infty}\frac{1}{\tau \tau^\prime}
\sum_{t\leq \tau}\sum_{t^\prime\leq
  \tau^\prime}C_{tt^\prime}]
\nonumber \\ &=& [1+\bra m_T^2\ket_\star]/(1+k)^{2}.
\label{eq:persistent_eta}
\end{eqnarray}
Note that $\bra
m_T^2\ket_\star=\lim_{\tau\to\infty}\tau^{-1}\sum_{t\leq
\tau}C(t)=c$.

The integrated response (or static susceptibility)
$k=\lim_{\tau\rightarrow\infty}\sum_{t\leq \tau} G(t)$ is also
calculated along the lines of \cite{HeimelCoolen}. One writes the
response function as $G_{tt^\prime}=\alpha^{-\frac{1}{2}}\bra
\partial~\sigma[q(t),z_t|T]/\partial \eta(t^\prime) \ket_\star$.
Integration by parts in this expression generates
\begin{equation}
 \bra
\partial~\sigma[q(t),z_t|T]/\partial \eta(t^\prime) \ket_\star
=\sum_{t^\pprime}\Sigma^{-1}_{t^\prime t^\pprime} \bra
\sigma[q(t),z_t|T] \eta(t^\pprime)\ket_\star,
\end{equation}
and
hence
\begin{equation}
\sqrt{\alpha}\sum_{t^\pprime}\bra \eta(t^\prime)\eta(t^\pprime)\ket
G^T_{t^\pprime t}
=
\bra \sigma[q(t),z_t|T]~ \eta(t^\prime)\ket_\star.
\label{eq:noise_relation}
\end{equation}
Averaging over the two times $t$ and $t^\prime$ now gives
in a stationary state without
anomalous response:
\begin{eqnarray}
\bra m_T \eta\ket_\star&=& k\sqrt{\alpha}\bra \eta^2 \ket.
\end{eqnarray}
Inserting the variance $\bra \eta^2\ket$, as given in
(\ref{eq:persistent_eta}), then gives
the general relation
\begin{eqnarray}
\bra \eta m_T \ket_\star&=& \frac{k\sqrt{\alpha}(1+c)}{(1+k)^{2}}.
\label{eq:eqn_for_k}
\end{eqnarray}

\subsection{Additive Decision Noise}

In the case of additive decision noise (\ref{eq:additive}) we have
$\sigma[q,z|T]=\sgn[q+zT]$. The effective agent is frozen if
$\widetilde{q}\not=0$, in which case $m_T= \sgn[
\widetilde{q}_T]$. This solves equation (\ref{eq:stationarity}) if
and only if $|\eta|>\sqrt{\alpha}/(1+k)$.  If
$|\eta|<\sqrt{\alpha}/(1+k)$, on the other hand, the
 agent is not frozen; now
$\widetilde{q}_T=0$ and $m_T=(1+k)\eta/\sqrt{\alpha}$. As a
result, we can calculate $c=\bra m_T^2\ket_\star$ and the fraction
$\phi=\bra\theta[
|\eta|-\sqrt{\alpha}/(1+k)]\ket=1-\erf[\sqrt{\alpha/2(1+c)}]$ of
frozen agents exactly as in in the case \cite{HeimelCoolen}
without decision noise, giving the deterministic (i.e.
$W(T)=\delta(T)$) result
\begin{equation}
  c=1-(1-\frac{1+c}{\alpha})~\erf\left[\sqrt{\frac{\alpha}{2(1+c)}}\right]
 -2\sqrt{\frac{1+c}{2\pi \alpha}}e^{-\frac{\alpha}{2(1+c)}}.
  \label{eq:additive_c}
\end{equation}
We use (\ref{eq:eqn_for_k}) and calculate the covariance $\bra
\eta m_T \ket_\star$ exactly as in \cite{HeimelCoolen}. The final
result is
\begin{equation}
\label{eq:additive_k}
 \frac{1}{k}=\frac{\alpha}{\erf[\sqrt{\frac{\alpha}{2(1+c)}}]}-1,
\end{equation}
with the value of $c$ to be determined by solving Eqn.
(\ref{eq:additive_c}). We find exactly the same transition point
$\alpha_c\approx 0.33740$,

\begin{figure}[t]
\setlength{\unitlength}{0.95mm} \hspace*{-5mm}
\begin{picture}(100,85)
\put(10,  10){\epsfysize=80\unitlength\epsfbox{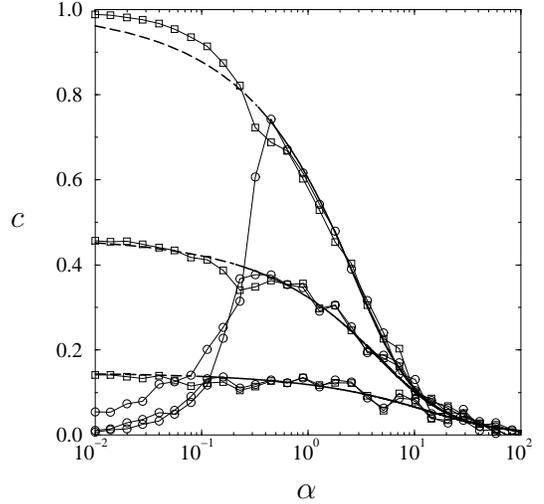}}
\put(11, 52){\large $c$} \put( 51,  14){\large $\alpha$}
\end{picture}
\vspace*{-12mm} \caption{The persistent correlation $c$ as a
function of  $\alpha=p/N$, for multiplicative noise with
$W(T)=\delta[T-\overline{T}]$ and different choices of the noise
strength ($\overline{T}=0,~1,~2$ from top to bottom). Connected
markers: individual simulation runs, with $pN=\alpha N^2=10^6$ and
homogeneous initial conditions where $q_i(0)=q(0)$ (circles:
$q(0)=0$, squares: $q(0)=10$) and in excess of 1000 iteration
steps. Thick solid curves for $\alpha>\alpha_c(W(T))$: analytical
predictions for homogeneous multiplicative decision noise. For
$\alpha<\alpha_c(W(T))$, where they should no longer be correct,
they have been continued as thick dashed lines. For additive
decision noise our theory predicts independence of $\overline{T}$
for $\alpha>\alpha_c(W(T))$, i.e. $c$ as given  by the
$\overline{T}=0$ curve of multiplicative noise.}
\label{fig:unif_correlation}
\end{figure}

\noindent
 signaling the divergence of the integrated response
$k$, as was found in the noise-free case.

Numerical simulations of the (batch) dynamics of the present model
(which we will not present here, for brevity) confirm quite convincingly
that, upon measuring objects such as $c$ or $\phi$,
in the case of additive decision noise one indeed exactly recovers the graphs of \cite{HeimelCoolen},
without any dependence on the noise parameters. This, however,
will turn out to be quite different in the case of multiplicative
noise.

\subsection{Homogeneous Multiplicative Decision Noise}

Next we turn to the case of multiplicative noise
(\ref{eq:multiplicative}), at first with the simplest distribution
$W(T)=\delta[T-\overline{T}]$, where
$\sigma[q,z|T]=\sgn[q]~\sgn[1+\overline{T}z]$, and where
$m_T=\lim_{\tau\to\infty}\tau^{-1}\sum_{t\leq
  \tau}\sgn[q_T(t)]~\sgn[1+\overline{T}z_t]$.
Since there is now only one noise strength in the system,
$\overline{T}$, we may drop the subscripts $T$ for variables such
as $q(t)$ or $m$, without danger of confusion.
  For a frozen agent one now finds
\begin{equation}
m=\lambda(\overline{T})~\sgn[\tilde{q}].
\end{equation}
This solves equation (\ref{eq:stationarity}) when
$|\eta|>\sqrt{\alpha}\lambda(\overline{T})/(1+k)$.  If
$|\eta|<\sqrt{\alpha}\lambda(\overline{T})/(1+k)$, on the other hand, the
 agent is

\begin{figure}[t]
\setlength{\unitlength}{0.95mm} \hspace*{-5mm}
\begin{picture}(100,85)
\put(10,
10){\epsfysize=80\unitlength\epsfbox{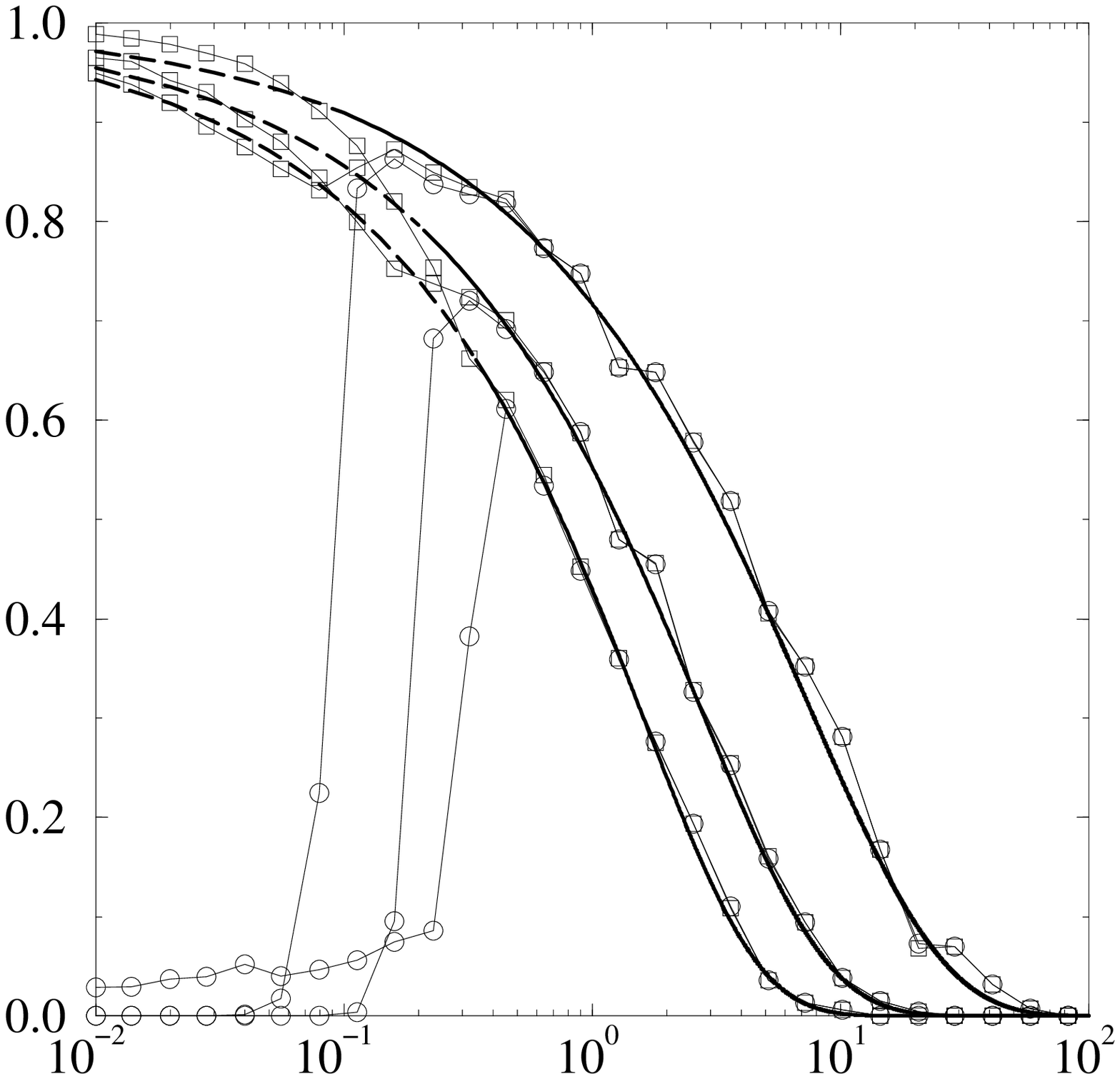}}
\put(11, 52){\large $\phi$} \put( 51,  14){\large $\alpha$}
\end{picture}
\vspace*{-12mm} \caption{The asymptotic fraction of frozen agents
$\phi$ as a function of  $\alpha=p/N$, for multiplicative noise
with $W(T)=\delta[T-\overline{T}]$ and different choices of the
noise strength ($\overline{T}=0,~1,~2$ from top to bottom).
Markers: individual simulation runs, with $pN=\alpha N^2=10^6$ and
homogeneous initial conditions where $q_i(0)=q(0)$ (circles:
$q(0)=0$, squares: $q(0)=10$) and in excess of 1000 iteration
steps. Thick solid curves for $\alpha>\alpha_c(W(T))$: analytical
predictions for homogeneous multiplicative decision noise. For
$\alpha<\alpha_c(W(T))$, where they should no longer be correct,
they have been continued as thick dashed lines. For additive
decision noise our theory predicts independence of $\overline{T}$,
i.e. $\phi$ as given  by the $\overline{T}=0$ curve of
multiplicative noise.} \label{fig:unif_frozen}
\end{figure}

\noindent not frozen; now $\widetilde{q}_T=0$ and
$m=(1+k)\eta/\sqrt{\alpha}$. We can again calculate $c=\bra
m^2\ket_\star$ self-consistently, upon distinguishing between the
two possibilities:
\begin{eqnarray}
c&=&\lambda^2(\overline{T})
\bra\theta\left[|\eta|-\frac{\sqrt{\alpha}\lambda(\overline{T})}{1+k}\right]
\ket
\\
&& +
\frac{(1+k)^2}{\alpha}\bra\theta\left[\frac{\sqrt{\alpha}\lambda(\overline{T})}{1+k}-|\eta|\right]\eta^2
\ket.
\end{eqnarray}
Working out the Gaussian integrals describing the statics of
$\eta$,
with variance (\ref{eq:persistent_eta}), subsequently gives
\begin{eqnarray}
  c&=&\lambda^2(\overline{T})-\left[\lambda^2(\overline{T})-\frac{1+c}{\alpha}\right]
  \erf\left[\sqrt{\frac{\alpha\lambda^2(\overline{T})}{2(1+c)}}\right]
\nonumber \\
 &&-2\lambda(\overline{T})\sqrt{\frac{1+c}{2\pi \alpha}}e^{-\frac{\alpha\lambda^2(\overline{T})}{2(1+c)}}.
  \label{eq:multiplicative_c}
\end{eqnarray}
 From this equation the value of $c$ is solved
numerically. The fraction $\phi$ of frozen agents is given by
\begin{equation}
\phi=\bra\theta\left[|\eta|-\frac{\sqrt{\alpha}\lambda(\overline{T})}{1+k}\right]
\ket
=1-\erf\left[\sqrt{\frac{\alpha\lambda^2(\overline{T})}{2(1+c)}}\right].
\label{eq:unif_phi}
\end{equation}
 We calculate the remaining object $\bra  \eta m\ket_\star$ in
(\ref{eq:eqn_for_k}) by again distinguishing between frozen and
non-frozen agents and

\begin{figure}[t]
\setlength{\unitlength}{0.95mm} \hspace*{-3mm}
\begin{picture}(100,85)
\put(10,
10){\epsfysize=80\unitlength\epsfbox{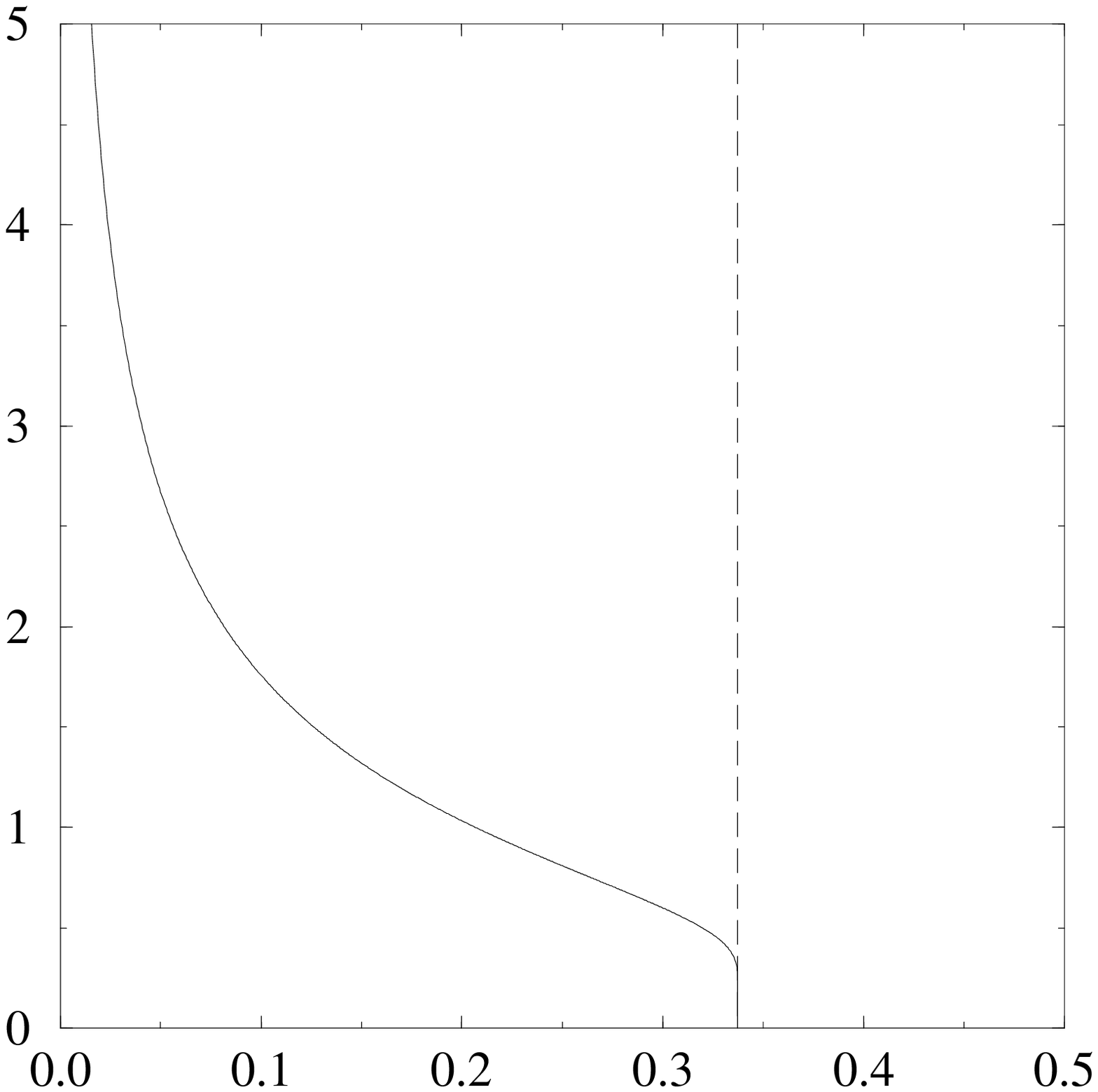}}
\put(11, 52){\large $\overline{T}$} \put( 51,  14){\large
$\alpha$}
\end{picture}
\vspace*{-12mm} \caption{ Phase diagram in the
($\alpha,1\minus\lambda(\overline{T})$) plane for homogeneous
multiplicative noise, i.e. $W(T)=\delta[T-\overline{T}]$. The
solid line separates a non-ergodic phase with anomalous response
(left) from an ergodic one without anomalous response (right). For
additive noise our theory predicts the $\overline{T}$-independent
transition given by the dashed line. } \label{fig:phasediag2}
\end{figure}

\noindent
 by using the two identities
$m=\lambda(T)\sgn[\eta]$ (for frozen agents) and
$m=\eta(1+k)/\sqrt{\alpha}$ (for fickle ones), both of which
follow from (\ref{eq:stationarity}), giving
\begin{eqnarray*}
\bra  \eta m\ket_\star&=&
\lambda(\overline{T})\bra\theta\left[|\eta|-\frac{\sqrt{\alpha}\lambda(\overline{T})}{1+k}\right]|\eta|\ket
~~~~~~~~
\\
&&
~~~~~~~~
+ \frac{1+k}{\sqrt{\alpha}} \bra\theta\left[\frac{\sqrt{\alpha}\lambda(\overline{T})}{1+k}-|\eta|\right]
\eta^2
\ket
\\
&=& \frac{1+c}{(1+k)\sqrt{\alpha}}~
\erf\left[\sqrt{\frac{\alpha\lambda^2(\overline{T})}{2(1+c)}}\right].
\end{eqnarray*}
Insertion into (\ref{eq:eqn_for_k}), together with
(\ref{eq:persistent_eta}), then gives the desired expression for
the integrated response:
\begin{equation}
\label{eq:multiplicative_k}
  \frac{1}{k}=\frac{\alpha}{\erf[\sqrt{\frac{\alpha\lambda^2(T)}{2(1+c)}}]}-1,
\end{equation}
 with the value of $c$ to be determined by solving Eqn.
(\ref{eq:multiplicative_c}).
Equivalently, using
(\ref{eq:unif_phi}) we find, as in the $T=0$ case
\cite{HeimelCoolen}
\begin{equation}
k=\frac{1-\phi}{\alpha-1+\phi}.
\label{eq:k2}
\end{equation}
 The integrated response $k$ is positive and finite, and
our solution exact, for $\alpha>\alpha_c(W(T))$. At
$\alpha_c(W(T))$ one finds that   $k$ diverges; this transition
is, as for $T=0$, found to happen when the fraction of fickle
agents equals $\alpha$ \cite{ChalMarsZecc00}. Finally, according
to (\ref{eq:multiplicative_c},\ref{eq:multiplicative_k}) we can
write $\alpha_c(W(T))$ as $\alpha_c(W(T))=\erf[x]$, where $x$ is
the solution of the transcendental equation
\begin{equation}
\lambda^2(\overline{T})\left\{\erf[x]-1+\frac{1}{x\sqrt{\pi}}e^{-x^2}\right\}=1.
\end{equation}
Equivalently, we can write our transition line explicitly in terms of the inverse error function as
\begin{equation}
\lambda(\overline{T}_c) =\left\{\alpha_c+\frac{e^{-\left[{\rm
erf}^{\rm inv}[\alpha_c]\right]^2}}{\erf^{\rm
inv}[\alpha_c]\sqrt{\pi}} -1\right\}^{-\frac{1}{2}},
\label{eq:phase_transition}
\end{equation}
where $\lambda(\overline{T})\in[0,1]$,  see (\ref{eq:lambdaT}).
\vsp

In figures \ref{fig:unif_correlation} and \ref{fig:unif_frozen} we
show the solution of equation (\ref{eq:multiplicative_c}) and the
corresponding fraction $\phi$ of frozen agents as functions of
$\alpha$, together with the values for $c$ and $\phi$ as obtained
by carrying out numerical simulations of the batch minority game
 (\ref{eq:batch}) with homogeneous multiplicative decision noise.
 The two figures for $c$ and $\phi$ both show excellent agreement
between theory and experiment above $\alpha_c(W(T))$. One observes
that, in addition to a reduction in the persistent correlation,
another  effect of the introduction of multiplicative decision
noise is an overall increase in the fraction of frozen agents.
This is consistent with our solution of the first few iteration
steps, where introducing decision noise had the effect of
dampening the oscillations. In figure \ref{fig:phasediag2} we show
the system's phase diagrams for $W(T)=\delta[T-\overline{T}]$,
defined by the transition line, where $k=\infty$. This line is
given by the solution of equation (\ref{eq:phase_transition}) in
the case of multiplicative noise, and by $\alpha_c(W(T))\approx
0.33740$ (i.e. the value corresponding to $\lambda(0)=1$) for
additive noise. Below $\alpha_c(W(T))$ our simulations show, as
has been observed and reported earlier for the deterministic case,
that in the anomalous response region the stationary state reached
by the system depends critically on the initial conditions. For
small values of the $|q_i(0)|$ (i.e. weak initial strategy
preferences) the system enters a high-volatility state with low
$c$ and $\phi$, whereas for large values of the $|q_i(0)|$ (i.e.
strong initial strategy preferences) the system enters a
low-volatility state with large $c$ and $\phi$.

\subsection{Inhomogeneous Multiplicative Decision Noise}

Finally we turn to the more complicated situation of
multiplicative noise (\ref{eq:multiplicative}) with arbitrary
distributions.
  For a frozen agent and for a given value of $T$ one has
\begin{equation}
m_T=\lambda(T)~\sgn[\tilde{q}].
\end{equation}
As before,
this solves equation (\ref{eq:stationarity}) if
$|\eta|>\sqrt{\alpha}\lambda(T)/(1+k)$, whereas for
$|\eta|<\sqrt{\alpha}\lambda(T)/(1+k)$ the
 agent is fickle, i.e.
$\widetilde{q}_T=0$ and $m_T=(1+k)\eta/\sqrt{\alpha}$. According
to (\ref{eq:finalC},\ref{eq:finalG}), the calculation of
persistent order parameters will now also involve averaging over
the noise distribution. Since the macroscopic dynamics turns out
to depend on $T$ only via $\lambda(T)$, it will be advantageous to
define
$w(\lambda)=\int_0^\infty\!\!dT~W(T)\delta[\lambda-\lambda(T)]$,
or
\begin{equation}
w(\lambda) =\int_0^\infty\!\!dT~W(T)~\delta\left[\lambda\!-\!
\int\!dz~P(z)~\sgn[1\!+\! Tz]\right].
\label{eq:define_wlambda}
\end{equation}

\begin{figure}[t]
\setlength{\unitlength}{0.95mm} \hspace*{-5mm}
\begin{picture}(100,85)
\put(10,  10){\epsfysize=80\unitlength\epsfbox{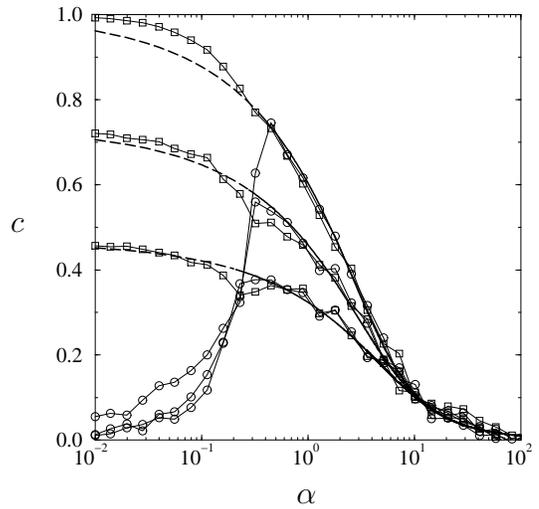}}
\put(11, 52){\large $c$} \put( 51,  14){\large $\alpha$}
\end{picture}
\vspace*{-12mm} \caption{The persistent correlation $c$ as a
function of  $\alpha=p/N$, for multiplicative noise with
$W(T^\prime)=\epsilon\delta[T^\prime-T]+(1-\epsilon)\delta[T^\prime]$,
for $T=1$ and different choices of the width ($\epsilon=0,~0.5,~1$
from top to bottom). Markers: individual simulation runs,
 with $pN=\alpha N^2=10^6$ and
homogeneous initial conditions where $q_i(0)=q(0)$ (circles:
$q(0)=0$, squares: $q(0)=10$) and in excess of 1000 iteration
steps.
 Solid
curves for $\alpha>\alpha_c(W(T))$: analytical predictions. For
$\alpha<\alpha_c(W(T))$, where they should no longer be correct,
they have been continued as dashed lines.}
\label{fig:bimodal_correlation}
\end{figure}

\noindent
 Here $\lambda\in[0,1]$, with $\lambda=0$ reflecting
$T\to\infty$ contributions and $\lambda=1$ reflecting $T\to 0$
ones. Now we may write
\begin{eqnarray}
c&=&\int_0^1\!d\lambda~w(\lambda)\left\{
\lambda^2
\bra\theta\left[|\eta|-\frac{\sqrt{\alpha}\lambda}{1+k}\right]
\ket
~~~~~~
\right.
\nonumber \\
&&
~~~~~~
\left.
+ \frac{(1+k)^2}{\alpha}
\bra\theta\left[\frac{\sqrt{\alpha}\lambda}{1+k}-|\eta|\right]\eta^2
\ket
\right\}
\nonumber \\
&=&\int_0^1\!d\lambda~w(\lambda)\left\{
\lambda^2
-2\lambda\sqrt{\frac{1+ c}{2\pi \alpha}}e^{-\frac{\alpha\lambda^2}{2(1+c)}}
\right.
~~~~~~
\nonumber \\
&&
\left.
~~~~~~
-\left[\lambda^2-\frac{1+ c}{\alpha}\right]
  \erf\left[\sqrt{\frac{\alpha\lambda^2}{2(1+ c)}}\right]
  \right\}.
  \label{eq:general_multiplicative_c}
\end{eqnarray}
From this equation the value of $c$ is solved numerically. The
fraction $\phi$ of frozen agents is given by
\begin{equation}
\phi
=1-\int_0^1\!d\lambda~w(\lambda)~\erf\left[\sqrt{\frac{\alpha\lambda^2}{2(1+c)}}\right].
\label{eq:general_multiplicative_phi}
\end{equation}
 We calculate the remaining object $\bra \eta m_T\ket_\star$
 in (\ref{eq:eqn_for_k}) by again distinguishing
between frozen and non-frozen agents and by using the two
identities $m_T=\lambda(T)\sgn[\eta]$ (for frozen agents) and
$m_T=\eta(1+k)/\sqrt{\alpha}$ (for the non-frozen ones), both of
which follow from (\ref{eq:stationarity}), giving
\begin{eqnarray*}
\bra \eta m_T \ket_\star&=&
\frac{1+c}{(1+k)\sqrt{\alpha}}\int_0^1\!d\lambda~w(\lambda)
~\erf\left[\sqrt{\frac{\alpha\lambda^2}{2(1+c)}}\right].
\end{eqnarray*}
Insertion into (\ref{eq:eqn_for_k}), together with
(\ref{eq:persistent_eta}), then gives the desired expression for
the integrated response:
\begin{equation}
\frac{1}{k}
=
\frac{\alpha}{ \int_0^1\!d\lambda~w(\lambda)
\erf\left[\sqrt{\frac{\alpha\lambda^2}{2(1+c)}}\right] }-1,
\label{eq:general_multiplicative_k}
\end{equation}
with the value of $c$ to be determined by solving Eqn.
(\ref{eq:general_multiplicative_c}). Using
(\ref{eq:general_multiplicative_phi})
this can again be written in the familiar form
(\ref{eq:k2}),
which suggests that the $k=\infty$ transition is of a geometrical
nature.

Unless we revert back to uniform noise levels, a transformation
like $\alpha_c(W(T))=\erf[x]$ will now no longer be helpful; to
find the location of the phase transition one has to solve
(\ref{eq:general_multiplicative_c}), together with the condition
$k=\infty$. Upon putting $y^2=\alpha/2(1+c)$ one can, however,
compactify these two coupled equations to
\begin{eqnarray}
1&=&\int_0^1\!d\lambda~w(\lambda)\lambda^2\left\{
\erf[y\lambda]-1+ \frac{e^{-y^2\lambda^2}}{y\lambda\sqrt{\pi}}
\right\}, \label{eq:general_transition1}
\\
\alpha&=& \int_0^1\!d\lambda~w(\lambda) \erf[y\lambda].
\label{eq:general_transition2}
\end{eqnarray}
\vsp

We will finally work out our equations describing the system with
inhomogeneous multiplicative decision noise explicitly for the
following simple bi-modal distribution
\begin{equation}
W(T^\prime)=\epsilon~\delta[T^\prime-T]+(1-\epsilon)\delta[T^\prime],
\label{eq:bimodal_W}
\end{equation}
with $\epsilon\in[0,1]$. For $\epsilon=1$ we revert back to the
homogeneous case studied earlier in this section; for $\epsilon=0$
we return to the model of \cite{HeimelCoolen}.
Here we have
\begin{eqnarray}
w(\lambda)&=&\epsilon\delta[\lambda-\lambda(T)]
+(1-\epsilon)\delta[\lambda-1],
\label{eq:bimodal_w}
\end{eqnarray}
with the function $\lambda(T)$ as defined in (\ref{eq:lambdaT}).
The general equations
(\ref{eq:general_multiplicative_c},\ref{eq:general_multiplicative_phi})
from which to solve $c$ and $\phi$ reduce to
\begin{eqnarray}
c
&=&
\epsilon\left\{
\lambda^2(T)
-2\lambda(T)
\sqrt{\frac{1\plus c}{2\pi \alpha}}e^{-\frac{\alpha\lambda^2(T)}{2(1+c)}}
\right.
\nonumber \\
&&
\left.~~~~~~~~
-\left[\lambda^2(T)-\frac{1\plus c}{\alpha}\right]
  \erf\left[\sqrt{\frac{\alpha\lambda^2(T)}{2(1+ c)}}\right]
\right\}
\nonumber \\
&&+
(1-\epsilon)\left\{
1-2\sqrt{\frac{1\plus c}{2\pi \alpha}}e^{-\frac{\alpha}{2(1+c)}}
\right.
\nonumber \\
&&
\left.~~~~~~~~~~~~~
-\left[1-\frac{1\plus c}{\alpha}\right]
  \erf\left[\sqrt{\frac{\alpha}{2(1+ c)}}\right]
\right\}, \label{eq:bimodal_c}
\end{eqnarray}
\begin{equation}
\phi=
1-\epsilon~\erf\left[\sqrt{\frac{\alpha\lambda^2(T)}{2(1+c)}}\right]
-(1-\epsilon) \erf\left[\sqrt{\frac{\alpha}{2(1+c)}}\right].
\label{eq:bimodal_phi}
\end{equation}
Similarly, the two coupled equations (\ref{eq:general_transition1},\ref{eq:general_transition2})
which define the phase transition reduce to

\begin{figure}[t]
\setlength{\unitlength}{0.95mm} \hspace*{-5mm}
\begin{picture}(100,85)
\put(10, 10){\epsfysize=80\unitlength\epsfbox{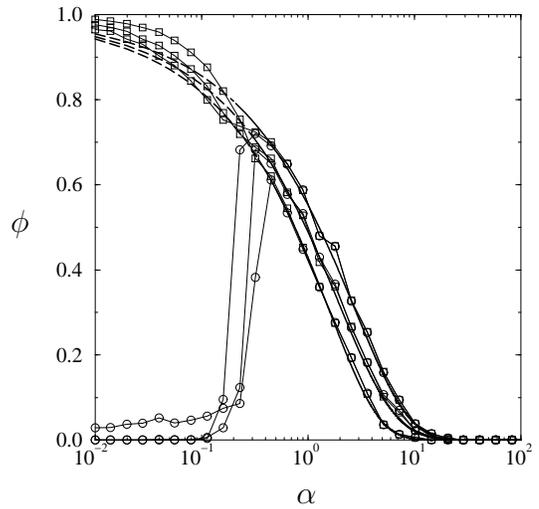}}
\put(11, 52){\large $\phi$} \put( 51,  14){\large $\alpha$}
\end{picture}
\vspace*{-12mm} \caption{The asymptotic fraction of frozen agents
$\phi$ as a function of  $\alpha=p/N$, for multiplicative noise
with
$W(T^\prime)=\epsilon\delta[T^\prime-T]+(1-\epsilon)\delta[T^\prime]$,
for $T=1$ and different choices of the width ($\epsilon=0,~0.5,~1$
from bottom to top). Markers: individual simulation runs,
 with $pN=\alpha N^2=10^6$ and
homogeneous initial conditions where $q_i(0)=q(0)$ (circles:
$q(0)=0$, squares: $q(0)=10$) and in excess of 1000 iteration
steps. Solid curves for $\alpha>\alpha_c(W(T))$: analytical
predictions. For $\alpha<\alpha_c(W(T))$, where they should no
longer be correct, they have been continued as dashed lines.}
\label{fig:bimodal_phi}
\end{figure}

\begin{eqnarray}
1 &=&\epsilon~\lambda^2(T) \left\{ \erf[y\lambda(T)]-1+
\frac{e^{-y^2\lambda^2(T)}}{y\lambda(T)\sqrt{\pi}} \right\}
\nonumber \\ && ~~~~~~~~ +(1-\epsilon) \left\{ \erf[y]-1+
\frac{e^{-y^2}}{y\sqrt{\pi}} \right\},
\label{eq:bimodal_transition1}
\\
\alpha&=& \epsilon~ \erf[y\lambda(T)] +(1-\epsilon)\erf[y].
\label{eq:bimodal_transition2}
\end{eqnarray}
Note that for $T\to 0$ our transition line equations reduce once
more to those of the noise-free case, as derived in
\cite{HeimelCoolen}, giving $\alpha_c\approx 0.33740$. For
$T\to\infty$, in contrast, we find a strong dependence on
$\epsilon$ (the fraction of traders who experience decision
noise). In particular, there is a qualitative difference between
$\epsilon<1$ and $\epsilon=1$ (where one of the two noise levels
in the system becomes zero).

 For $\epsilon=1$ we return to the case of uniform decision noise, and
equations
(\ref{eq:bimodal_transition1},\ref{eq:bimodal_transition2})
dictate that the transition line obeys $\alpha\to 0$ as $T\to
\infty$. For $\epsilon<1$ (i.e. a nonzero fraction of the traders
take decisions deterministically),
 on the other hand, we find for $T\to \infty$ the
equations
(\ref{eq:bimodal_transition1},\ref{eq:bimodal_transition2}) (which
will now have a solution with finite $y$) reducing to
\begin{eqnarray}
1 &=&(1-\epsilon) \left\{ \erf[y]-1+ \frac{e^{-y^2}}{y\sqrt{\pi}}
\right\},
\\
\alpha&=&(1-\epsilon)\erf[y].
\end{eqnarray}
Equivalently:
\begin{equation}
\sqrt{\pi}\left[\frac{2-\epsilon-\alpha}{1-\epsilon}\right]
\erf^{\rm inv}\left[\frac{\alpha}{1-\epsilon}\right]
=
e^{-\left[\erf^{\rm
inv}\left[\frac{\alpha}{1-\epsilon}\right]\right]^2}.
\end{equation}

\begin{figure}[t]
\setlength{\unitlength}{0.95mm} \hspace*{-3mm}
\begin{picture}(100,85)
\put(10,
10){\epsfysize=80\unitlength\epsfbox{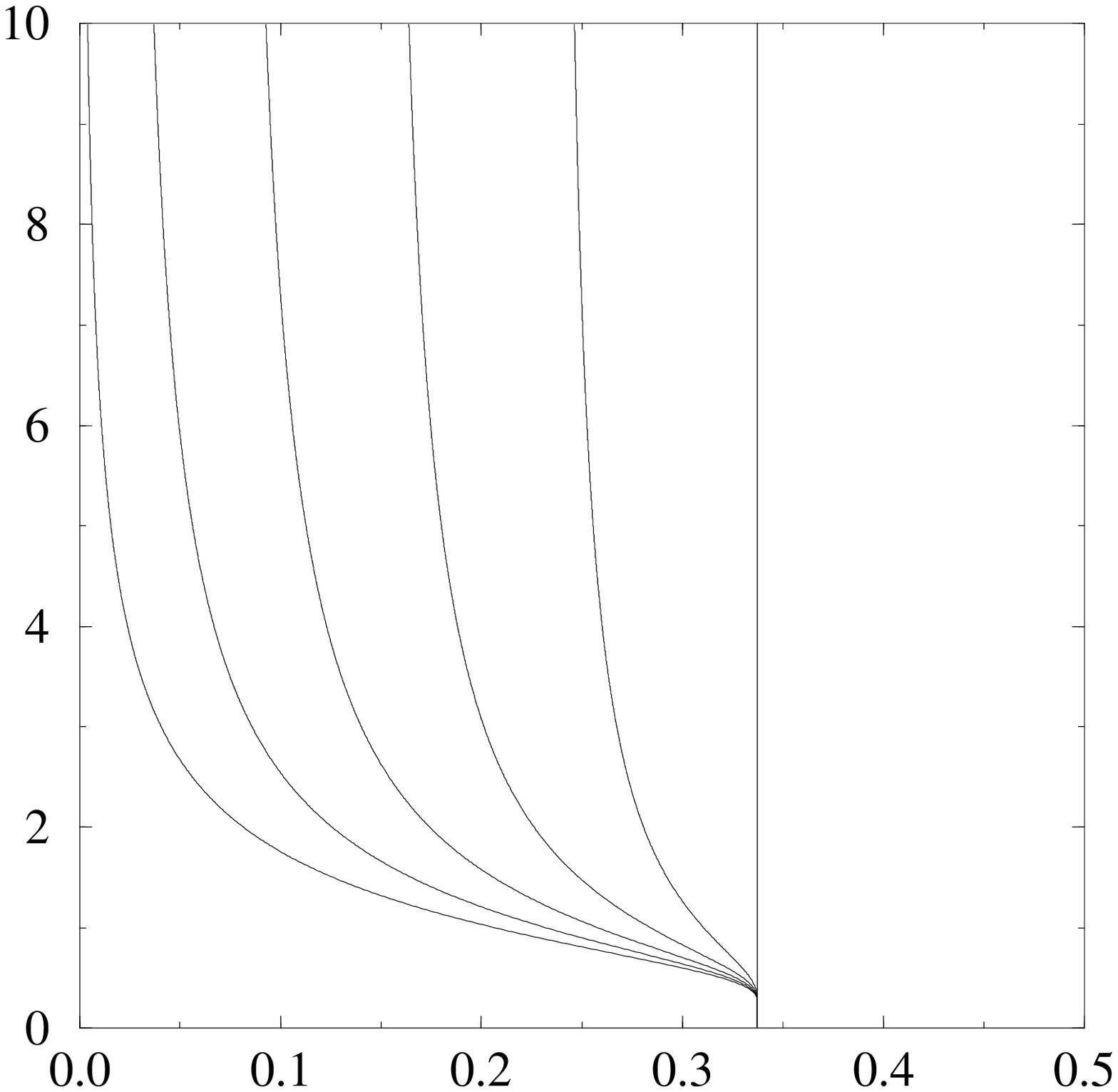}}
\put(11, 52){\large $T$} \put( 51,  14){\large $\alpha$}
\end{picture}
\vspace*{-12mm} \caption{The phase diagram for multiplicative
noise with
$W(T^\prime)=\epsilon\delta[T^\prime-T]+(1-\epsilon)\delta[T^\prime]$
and Gaussian distributed $z$,
 shown in the $(\alpha,T)$ plane for different values of $\epsilon$ ($\epsilon\in\{0,0.2,0.4,0.6,0.8,1\}$,
 from right to left).
For each value of $\epsilon$, the solid line separates a
non-ergodic phase with anomalous response (left) from an ergodic
one without anomalous response (right). For additive noise our
theory predicts the $T$-independent transition given by the
vertical line (i.e. the $\epsilon=0$ curve). }
\label{fig:bimodal_phasediag}
\end{figure}

\noindent
 The solution of this equation defines the point
$\alpha_c(\epsilon,T=\infty)$, which obeys
$\alpha_c(\epsilon<1,T=\infty)>0$ and $\alpha_c(1,T=\infty)=0$.

In figures \ref{fig:bimodal_correlation} and \ref{fig:bimodal_phi}
we show the (numerical) solution of equation (\ref{eq:bimodal_c})
for the persistent correlation $c$,  and the corresponding value
for the fraction $\phi$ of frozen agents, as given by
(\ref{eq:bimodal_phi}), as functions of $\alpha$ and for different
choices of the parameters $\{T,\epsilon\}$,
 together with the
corresponding values for $c$ and $\phi$ as obtained by carrying
out numerical simulations. Here we have chosen Gaussian
distributed $z_j(\ell)$, i.e. $\lambda(T)=\erf[1/T\sqrt{2}]$. As
before, one observes excellent agreement between theory and
experiment above $\alpha_c$, and a strong dependence on initial
conditions below $\alpha_c$. Finally, in figure
\ref{fig:bimodal_phasediag}  we show, in the
$(\alpha,\overline{T})$ plane, the system's phase diagram as
defined by the $k=\infty$ transition line, obtained by solving
numerically the coupled equations
(\ref{eq:bimodal_transition1},\ref{eq:bimodal_transition2}), for
different values of $\epsilon$.

\section{Stationary Volatility for $\alpha>\alpha_c(W(T))$}

As in the noise-free case \cite{HeimelCoolen}, one finds that the volatility matrix
(\ref{eq:volatility_matrix}), which is to be calculated from expressions (\ref{eq:noise_covariance})
and which in a stationary state is time-translation-invariant
$\Xi_{tt^\prime}=\Xi(t-t^\prime)$,
generally involves both
long-term and short-term fluctuations.
Hence even the ordinary single-time stationary volatility $\sigma^2=\Xi(0)$
cannot be expressed in terms of the persistent order parameter $c$
(or its relatives $k$ and $\phi$).
Upon separating in the functions $C$ and $G$ the persistent from the
non-persistent terms, i.e. $C(t)=c+\tilde{C}(t)$ and
$G(t)=\tilde{G}(t)$ (there is no persistent response for $\alpha>\alpha_c$),
we find, as in \cite{HeimelCoolen}:
\begin{eqnarray}
\sigma^2 &=& \frac{1+c}{2(1+k)^{2}}+ \nonumber \\ &&\hspace*{-5mm}
\lim_{\tau\to\infty}\frac{1}{2\tau}\sum_{u\leq\tau} \sum_{t^\prime
t^\pprime} (\openone+\tilde{G})_{u
t^\prime}^{-1}\tilde{C}_{t^\prime
t^\pprime}(\openone+\tilde{G}^T)_{t^\pprime u}^{-1}.
\label{eq:sigma2}
\end{eqnarray}
 Obtaining an exact expression for $\sigma^2$ would require
solving our coupled saddle-point equations
(\ref{eq:finalC},\ref{eq:finalG}) for $C_{tt^\prime}$ and
$G_{tt^\prime}$ for large times but finite temporal separations
$t-t^\prime$, hence in practice one has to resort to
approximations. The approximation chosen in
\cite{ChalMarsZecc00,MarsChalZecc00}, for instance, is in our
language equivalent to substituting $\bra \sigma[q_i(t),z_i(t)|T]
\sigma[q_j(t),z_j(t)|T]\ket \to \delta_{ij}+ (1\minus
\delta_{ij})\bra \sigma[q_i(t),z_i(t)|T]\ket\bra
\sigma[q_j(t),z_j(t)|T]\ket$. Here we will generalise to the case
of decision noise the (at least for the batch MG) slightly more
accurate approximation proposed in \cite{HeimelCoolen}. We will
abbreviate the double averages $\int\!dT~W(T)\bra \ldots\ket$ as
$\bra\bra \ldots\ket\ket$. In order to find the volatility we
separate the correlations at stationarity in a `frozen' and a
`fickle' contribution:
\begin{eqnarray}
C(t-t^\prime)&=&
\phi \bra\bra \sigma[q(t),z_t|T]\sigma[q(t^\prime),z_{t^\prime}|T]\ket\ket_{\rm fr}
\\
&& + (1-\phi)\bra\bra
\sigma[q(t),z_t|T]\sigma[q(t^\prime),z_{t^\prime}|T]\ket\ket_{\rm
fi},
\end{eqnarray}
which gives, using $\tilde{C}(t-t^\prime)=C(t-t^\prime)-c$, and
upon rewriting the `fickle' contribution to the volatility:
\begin{eqnarray}
\sigma^2 &=& \frac{1}{2(1+k)^{2}} \nonumber \\ &&\hspace*{-5mm}
+\lim_{\tau\to\infty}\frac{1-\phi}{2\tau}\sum_{u\leq\tau}\bra\bra
\left[\sum_{t} (\openone+\tilde{G})_{u
t}^{-1}\sigma[q(t),z_{t}|T]\right]^2 \ket\ket_{\rm fi} \nonumber\\
&& +\lim_{\tau\to\infty}\frac{\phi}{2\tau}\sum_{u\leq\tau} \sum_{t
t^\prime} (\openone+\tilde{G})_{u
t}^{-1}(\openone+\tilde{G}^T)_{t^\prime u}^{-1} \nonumber\\
&&\hspace*{10mm}\times \bra\bra
\sigma[q(t),z_{t}|T]\sigma[q(t^\prime),z_{t^\prime}|T]\ket\ket_{\rm
fr}.
\label{eq:sigma3}
\end{eqnarray}
The approximation of \cite{HeimelCoolen} consists of retaining in the contribution from
`fickle' agents only the instantaneous $u=t$ terms, the rationale being that
the $u\neq t$ ones represent in the original single-trader equation a retarded
self-interaction, which is assumed to be significant only for
`frozen' agents.
Hence we obtain
\begin{eqnarray}
\sigma^2 &=& \frac{1}{2(1+k)^{2}}+\frac{1}{2}(1-\phi) \nonumber\\
&& +\lim_{\tau\to\infty}\frac{\phi}{2\tau}\sum_{u\leq\tau} \sum_{t
t^\prime} (\openone+\tilde{G})_{u
t}^{-1}(\openone+\tilde{G}^T)_{t^\prime u}^{-1} \nonumber\\
&&\hspace*{10mm}\times \bra\bra
\sigma[q(t),z_{t}|T]\sigma[q(t^\prime),z_{t^\prime}|T]\ket\ket_{\rm
fr}.
\label{eq:sigma4}
\end{eqnarray}
Note that, according to
(\ref{eq:general_multiplicative_phi},\ref{eq:general_multiplicative_k}),
 the integrated response $k$ can be expressed in terms of the order parameter
$\phi$ as $k=(1-\phi)/(\alpha-1+\phi)$.
\vsp

At this stage we again have to distinguish between additive noise
and multiplicative noise, in order to work

\begin{figure}[t]
\setlength{\unitlength}{0.95mm} \hspace*{-5mm}
\begin{picture}(100,85)
\put(10,  10){\epsfysize=80\unitlength\epsfbox{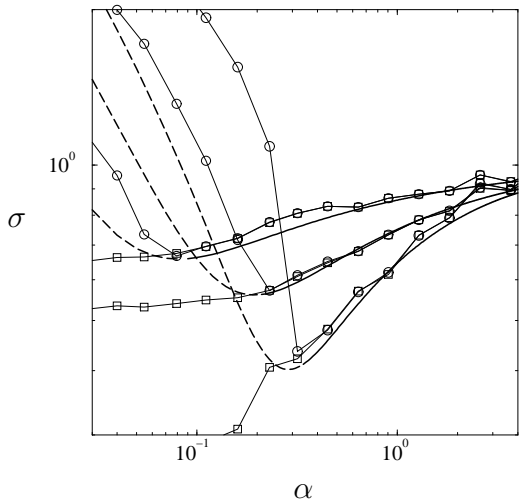}}
\put(11, 52){\large $\sigma$} \put( 51,  14){\large $\alpha$}
\end{picture}
\vspace*{-12mm} \caption{The asymptotic volatility $\sigma$ as a
function of  $\alpha$, for multiplicative noise with
$W(T)=\delta[T-\overline{T}]$ and different choices of the noise
strength ($\overline{T}=0,~1,~2$ from bottom to top in
$\alpha>\alpha_c$ regime). Markers: individual simulation runs,
with $pN=\alpha N^2=10^6$ and homogeneous initial conditions where
$q_i(0)=q(0)$ (circles: $q(0)=0$, squares: $q(0)=10$) and in
excess of 1000 iteration steps. Thick solid curves for
$\alpha>\alpha_c(W(T))$: analytical predictions for homogeneous
multiplicative decision noise. For $\alpha<\alpha_c(W(T))$, where
they should no longer be correct, they have been continued as
thick dashed lines. For additive decision noise our theory
predicts independence of $\overline{T}$, i.e. $\sigma$ as given by
the $\overline{T}=0$ curve of multiplicative noise.}
\label{fig:sigma_hom}
\end{figure}

\noindent
 out the remaining averages. For additive noise one
simply finds
\[
\bra\bra \sigma[q(t),z_{t}|T]\sigma[q(t^\prime),z_{t^\prime}|T]\ket\ket_{\rm fr}
=
\bra\bra
\sigma[\tilde{q}(t)]\sigma[\tilde{q}(t^\prime)]\ket\ket_{\rm
fr}=1,
\]
and hence we recover the expression describing the noise-free case
in \cite{HeimelCoolen}:
\begin{equation}
\sigma^2 =\frac{1+\phi}{2(1+k)^{2}}+\frac{1}{2}(1-\phi).
\label{eq:sigma5}
\end{equation}
Since the order parameters $\phi$ and $k$ are, for additive noise,
independent of the noise distribution, the same is true for the
volatility. This independence of the noise parameters, at least
for $\alpha>\alpha_c$ (in line with
\cite{ChalletPRL2000,CavagnaPRL2000}), again finds confirmation in
numerical simulations (that is, within the limits imposed by our
approximation; one does observe some weak effect, which could
either be due to excessive
 relation times or due to the retarded self-interaction
of `fickle' traders, which we neglected in deriving
(\ref{eq:sigma5})). \vsp

The more interesting case, as before, is that of multiplicative
noise. Here we have
\begin{eqnarray}
\bra\bra
\sigma[q(t),z_{t}|T]\sigma[q(t^\prime),z_{t^\prime}|T]\ket\ket_{\rm
fr} &=&\bra\bra \lambda^2(T)\ket\ket_{\rm fr} \nonumber\\[1mm]
&&\hspace*{-10mm} +\delta_{tt^\prime}[1-\bra\bra
\lambda^2(T)\ket\ket_{\rm fr}].
\end{eqnarray}
Hence the approximation (\ref{eq:sigma4}) reduces to

\begin{figure}[t]
\setlength{\unitlength}{0.95mm} \hspace*{-5mm}
\begin{picture}(100,85)
\put(10, 10){\epsfysize=80\unitlength\epsfbox{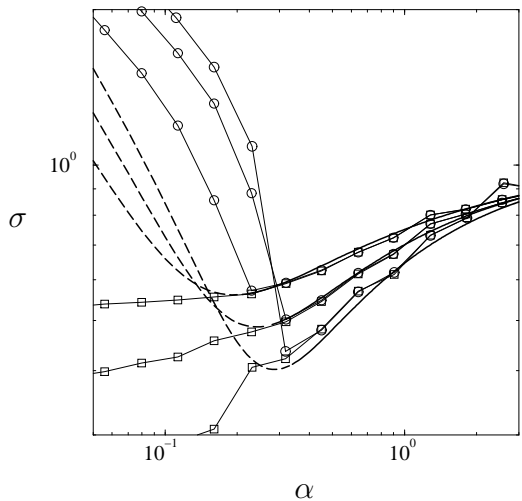}}
\put(11, 52){\large $\sigma$} \put( 51,  14){\large $\alpha$}
\end{picture}
\vspace*{-12mm} \caption{The asymptotic volatility $\sigma$ as a
function of $\alpha$, for multiplicative noise with
$W(T^\prime)=\epsilon\delta[T^\prime-T]+(1-\epsilon)\delta[T^\prime]$,
for $T=1$ and different choices of the width ($\epsilon=0,~0.5,~1$
from bottom to top in the $\alpha>\alpha_c$ regime). Markers:
individual simulation runs,
 with $pN=\alpha N^2=10^6$ and
homogeneous initial conditions where $q_i(0)=q(0)$ (circles:
$q(0)=0$, squares: $q(0)=10$) and in excess of 1000 iteration
steps. Solid curves for $\alpha>\alpha_c(W(T))$: analytical
predictions. For $\alpha<\alpha_c(W(T))$, where they should no
longer be correct, they have been continued as dashed lines.}
\label{fig:sigma_inhom}
\end{figure}

\begin{eqnarray}
\sigma^2 &=& \frac{1+\phi\chi}{2(1+k)^{2}}+\frac{1}{2}(1-\phi)
\nonumber\\ &&
+\frac{1}{2}\phi(1-\chi)[(\openone+\tilde{G})^{-1}(\openone+\tilde{G}^T)^{-1}](0).
\label{eq:sigma6}
\end{eqnarray}
Here we have used time-translation invariance of the stationary
state, giving $[\ldots]_{tt}\to [\ldots](t-t)=[\ldots](0)$ for the
relevant matrix elements in (\ref{eq:sigma6}).
 The conditional average $\chi=\bra\bra
\lambda^2(T)\ket\ket_{\rm fr}$, constrained by
$|\eta|>\sqrt{\alpha}\lambda(T)/(1+k)$ (which, in the case of
multiplicative noise, is the condition for an agent to be frozen)
and calculated using the variance $\bra \eta^2 \ket =
(1+c)/(1+k)^{2}$ (\ref{eq:persistent_eta}) of the zero-average
persistent noise term, is given by
\begin{eqnarray}
\chi&=& \bra\bra \lambda^2(T)\ket\ket_{\rm fr} \nonumber\\ &=&
\frac{\int_0^\infty\!dT~W(T)\lambda^2(T)\int\!Dz~\theta\left[|z|-\frac{\sqrt{\alpha}\lambda(T)}{\sqrt{1+c}}\right]}
{\int_0^\infty\!dT~W(T)\int\!Dz~\theta\left[|z|-\frac{\sqrt{\alpha}\lambda(T)}{\sqrt{1+c}}\right]}
\nonumber
\end{eqnarray}

\noindent
\begin{eqnarray}
&=&
\frac{\int_0^1\!d\lambda~w(\lambda)\lambda^2\left[1-\erf\left(\frac{\lambda\sqrt{\alpha}}{\sqrt{2(1+c)}}\right)\right]}
{\int_0^1\!d\lambda~w(\lambda)\left[1-\erf\left(\frac{\lambda\sqrt{\alpha}}{\sqrt{2(1+c)}}\right)\right]}.
\label{eq:chi}
\end{eqnarray}
We note that only for $W(T)=\delta(T)$ \cite{HeimelCoolen}, i.e.
$w(\lambda)=\delta(\lambda-1)$, where $\chi=1$, will (\ref{eq:sigma6}) involve only persistent observables.
In the presence of
decision noise, as in this study, one always has $\chi<1$, and additional
approximations are required to reduce also the last term in (\ref{eq:sigma6}) further to an
expression in terms of persistent order parameters only.
This is done in detail in the appendix, where we show that a
reasonable approximation is obtained by simply putting
$[(\openone+\tilde{G})^{-1}(\openone+\tilde{G}^T)^{-1}](0)\to 1$.
The end result is the following final approximation
for the stationary state volatility:
\begin{eqnarray}
\sigma^2 &=& \frac{1+\phi\chi}{2(1+k)^{2}}+\frac{1}{2}(1-\phi)
+\frac{1}{2}\phi(1-\chi), \label{eq:sigma7}
\end{eqnarray}
with $\chi$ as given by (\ref{eq:chi}).
\vsp

Expression (\ref{eq:sigma7}), which reverts back to that of
\cite{HeimelCoolen} for $T\to 0$ and which also reduces correctly
to the random trading limit $\sigma=1$ for $T\to\infty$ (where
$\phi=1$, $c=k=\chi=0$), turns out to be a surprisingly accurate
approximation  of the volatility for  $\alpha>\alpha_c$ (i.e. in
its

\noindent
regime of validity). This can be observed in Figures
\ref{fig:sigma_hom} and \ref{fig:sigma_inhom}, where we compare
the approximate prediction (\ref{eq:sigma7}) to the volatility as
observed in numerical simulations, for both homogeneous
multiplicative noise defined by $W(T)=\delta[T-\overline{T}]$ and
for inhomogeneous multiplicative noise defined by
(\ref{eq:bimodal_W},\ref{eq:bimodal_w}), respectively. In all
cases $\lambda(T)=\erf[1/T\sqrt{2}]$ (note: the persistent order
parameters have already been calculated in the previous section).

The above results emphasize once more the qualitative difference
between additive and multiplicative noise: in contrast to additive
noise, the system remains sensitive to multiplicative noise even
for $\alpha>\alpha_c$. The resulting dependence of the volatility
on the multiplicative noise strength is very similar to that
reported in \cite{CavaGarrGiarSher99} for additive noise (which
was later understood to be caused by insufficient equilibration
\cite{ChalletPRL2000,CavagnaPRL2000}).

\section{Discussion}

In this paper we have generalised the Thermal Minority Game
\cite{CavaGarrGiarSher99} to the case of imhomogeneous agent
populations (where the decision noise, which can be either
additive or multiplicative, is of non-uniform strength). We have
solved the dynamics of the batch version of this model by
generalizing the recent application \cite{HeimelCoolen} to the
Minority Game of the generating functional techniques of
\cite{DeDominicis} (note that in \cite{HeimelCoolen} only the
fully deterministic case was studied). This formalism reduces the
$N$-agent dynamics, in the limit $N\to\infty$, to a stochastic
process for a single `effective agent', with dynamic equations
involving coloured noise and a retarded self-interaction. It leads
to exact closed (but implicit and non-trivial) equations for
correlation- and response-functions.

 Our theory enables us to (i) obtain an analytical and quantitative
understanding of previously observed but unexplained phenomena
(e.g. suppression of the volatility by decision noise, even below
random, due to damping of the `crowd anti-crowd' oscillations),
(ii) derive exact phase diagrams, and (iii) calculate macroscopic
observables (e.g. the fraction of frozen agents and the persistent
correlations) in ergodic stationary states
exactly\footnote{Although the stationary state  equations derived
upon assuming ergodicity and absence of long-term memory are no
longer valid in the non-ergodic regime,
 Figures 1,2, 4 and 5 show that for $\alpha<\alpha_c(W(T))$ their
 predictions regarding the persistent observables $c$ and $\phi$
 nevertheless give good qualitative agreement with the results of
 simulations from a highly biased start (for the volatility
 $\sigma$, which also involves non-persistent order parameters,
 this is no longer the case).}.
 In the case of additive decision noise we find
a phase diagram identical to that of deterministic decision making
in the onset of and equilibrium properties of the higher $\alpha$
ergodic phase, with non-ergodic behaviour at lower $\alpha$. In
the case of multiplicative decision noise, in contrast, we arrive
at phase diagrams with non-trivial decision noise dependencies of
the phase separation line as well as the behaviour of both phases.
Here the control parameters are the relative number of possible
value for the external information, $\alpha=p/N$, and the
parameters characterizing the noise statistics. In the non-ergodic
regime of the model (i.e. for sufficiently small $\alpha$), our
closed equations in terms of correlation- and response functions
are still exact, and can be solved in principle iteratively for
arbitrary times; however, finding the stationary states is hard
(see e.g. the calculations for the simpler case
\cite{HeimelCoolen})\footnote{Note that a recently proposed
procedure \cite{HeimelDM} for calculating at least the
high-volatility stationary state in the non-ergodic regime, based
on assuming the integrated response function (which diverges
exactly at the critical point) to remain infinite throughout the
$\alpha<\alpha_c$ region, is not likely to work for the case of
decision noise. It would, for instance, predict the simple
relation $\phi=1-\alpha$ (i.e. $\phi$ being independent of the
noise parameters), which is clearly in conflict with the
simulation experiments presented in this paper.}. Here we have
restricted our calculations in the non-ergodic regime to the the
first few time-steps, finding noise dependence for both additive
and multiplicative decision noise.

In the present paper we have only worked out explicitly two types
of choices for the decision noise strengths statistics: a delta
distribution (i.e. decision noise of uniform strength), and a
parametrized class of bi-model distributions. Due to the general
nature of our solution, however, there is no limit to the
different types of noise statistics we could have studied. This
emphasizes once more the remarkable potential and appropriateness
to the Minority Games of the generating functional analysis
methods of \cite{DeDominicis}. Two natural next steps would be  to
develop the generating functional formalism for the original
`on-line' formulation of the game, where the external information
is fed to the agents sequentially (this is the subject of
\cite{Coolen}), or to analyze our present (exact) order parameter
equations further in the non-ergodic region
$\alpha<\alpha_c(W(T))$.

\appendix
\section{Approximation of Non-Persistent Terms in the Stationary
Volatility}

The term $Q=[(\openone+\tilde{G})^{-1}(\openone+\tilde{G}^T)^{-1}](0)$
in (\ref{eq:sigma6}), which contains contributions of
non-persistent order parameters,
can be written as
\begin{equation}
Q=
\int_{-\pi}^\pi\!\frac{d\omega}{2\pi}\frac{1}{|1+\hat{G}(\omega)|^2},
\label{eq:nastyterm1}
\end{equation}
with the definition
$\hat{G}(\omega)=\sum_t \tilde{G}(t)e^{-i\omega t}$.
The simplest approximation for $\tilde{G}(t)$, which
respects causality and also meets
the requirement $\sum_t \tilde{G}(t)=k$,  is an exponential expression of the form
$\tilde{G}(t>0)\to k(1-\gamma)\gamma^{t-1}$ (with $-1<\gamma<1$ and with $\tilde{G}(t\leq 0)=0$).
This gives
$\hat{G}(\omega)=k(1-\gamma)/(e^{i\omega}-\gamma)$,
and thus
\begin{eqnarray}
Q &=&
\int_{-\pi}^\pi\!\frac{d\omega}{2\pi}\frac{|e^{i\omega}-\gamma|^2}{|e^{i\omega}-\gamma+k(1-\gamma)|^2}
\label{eq:fourier}
\end{eqnarray}
We will obtain an estimate for $\gamma$ by carrying out an
approximate calculation of the one-step response function
\begin{equation}
\tilde{G}(1)=\frac{\partial }{\partial\theta({t})} \bra~
\sigma[q(t+1),z_{t+1}|T]~\ket_\star
\end{equation}
We insert (\ref{eq:singleagent}), we use the fact that the
response of frozen agents will be zero, we repeat our previous
ansatz that fickle agents do not experience a retarded
self-interaction, and we carry out the average over the decision
noise variable $z_t$. This is followed by carrying out the average
over $\eta(t)$ (which is Gaussian, with variance $\bra
\eta^2(t)\ket =2\sigma^2$; we assume, within the context of the
present approximation, the correlations between $\eta(t)$ and the
persistent noise $\eta$ not to be important for fickle agents).
This gives
\begin{eqnarray}
\tilde{G}(1)
&=&\frac{1-\phi}{\sigma\sqrt{\pi\alpha}}~
\bra\bra \lambda(T)e^{-[q^2(t)/\alpha+\alpha]/4\sigma^2}
\nonumber
\\
&&\times \left[ \cosh[\frac{|q(t)|}{2\sigma^2}]
+\lambda(T)\sinh[\frac{|q(t)|}{2\sigma^2}] \right] \ket\ket_{\rm
fi}.
\label{eq:G1_a}
\end{eqnarray}
In this expression we simply replace $|q(t)|\to 0$ (fickle agents
being described by values of $q(t)$ which oscillate around zero), and we
calculate the residual average $\bra\bra \lambda(T)\ket\ket_{\rm fi}$
similar to our calculation of (\ref{eq:chi}).
Hence we arrive at the approximation
\begin{equation}
\tilde{G}(1)\approx \frac{1-\phi}{\sigma\sqrt{\pi\alpha}}
e^{-\alpha/4\sigma^2}\left\{
\frac{\int_0^1\!d\lambda~w(\lambda)\lambda
\erf\left(\frac{\lambda\sqrt{\alpha}}{\sqrt{2(1+c)}}\right)}
{\int_0^1\!d\lambda~w(\lambda)\erf\left(\frac{\lambda\sqrt{\alpha}}{\sqrt{2(1+c)}}\right)}
\right\}.
\label{eq:G1_b}
\end{equation}
On the other hand, according to our ansatz $\tilde{G}(t>0)= k(1-\gamma)\gamma^{t-1}$
we must demand
$\tilde{G}(1)= k(1-\gamma)$, so that (\ref{eq:G1_b}) leads to the
following estimate of $\gamma$:
\begin{equation}
\gamma\approx 1-\frac{1-\phi}{\sigma k\sqrt{\pi\alpha}}
e^{-\alpha/4\sigma^2}\left\{
\frac{\int_0^1\!d\lambda~w(\lambda)\lambda
\erf\left(\frac{\lambda\sqrt{\alpha}}{\sqrt{2(1+c)}}\right)}
{\int_0^1\!d\lambda~w(\lambda)\erf\left(\frac{\lambda\sqrt{\alpha}}{\sqrt{2(1+c)}}\right)}
\right\}.
\label{eq:gamma}
\end{equation}
Since for $\alpha\to \infty $ we must find $\sigma\to 1$ (random
trading), and since $k\sim \alpha^{-1}$ (\ref{eq:k2}), we conclude
from (\ref{eq:gamma}) that $\gamma\to 1$ for $\alpha\to \infty$.
Conversely, as $\alpha$ is lowered, we find a divergence of $k$ at
finite $\alpha_c$ (where also $\phi$ is finite). Hence
(\ref{eq:gamma}) also predicts that $\gamma\to 1$ for $\alpha\to
\alpha_c$. We now assume that $\gamma\to 1$ will give a sensible
approximation in the whole range $\alpha>\alpha_c$, and use
(\ref{eq:fourier}) to arrive at the approximate result
\begin{equation}
[(\openone+\tilde{G})^{-1}(\openone+\tilde{G}^T)^{-1}](0)\approx
1.
\label{eq:non_persistent}
\end{equation}
The above derivation is clearly far from rigorous, and not quite
satisfactory; it simply appears the best one can do without
actually solving the order parameter equations for finite time
differences in the stationary state.  Yet
(\ref{eq:non_persistent}) turns out to lead to a surprisingly
accurate approximation for the volatility (see the main text).


\begin{references}

\bibitem{ChalZhan97}
D. Challet and Y.-C. Zhang, Physica A {\bf 246},  407  (1997).


\bibitem{Savit}
R. Savit, R. Manuca and R. Riolo, Phys. Rev. Lett. {\bf 82}, 2203
(1999).

\bibitem{Arth94}
W. Arthur, Am. Econ. Assoc. Papers and Proc. {\bf 84},  406  (1994).

\bibitem{Chalweb}
D. Challet, http:/$\!$/www.unifr.ch/econophysics/minority/ (an extensive commented
  collection of work on the minority game).

\bibitem{Cava99}
A. Cavagna, Phys. Rev. E {\bf 59},  R3783  (1999).

\bibitem{CavaGarrGiarSher99}
A. Cavagna, J. Garrahan, I. Giardina and D. Sherrington, Phys.
Rev. Lett. {\bf
  83},  4429  (1999).

\bibitem{ChalMarsZecc00}
D. Challet, M. Marsili, and R. Zecchina, Phys. Rev. Lett. {\bf 84},  1824
  (2000).

\bibitem{MarsChalZecc00}
M. Marsili, D. Challet, and R. Zecchina, Physica A {\bf 280},  522  (2000).

\bibitem{GarrMoroSher00}
J. Garrahan, E. Moro, and D. Sherrington, Phys. Rev. E {\bf 62},  R9  (2000).

\bibitem{ChalletPRL2000}
D. Challet, M. Marsili and R. Zecchina, Phys. Rev. Lett. {\bf 85},
5008 (2000).

\bibitem{CavagnaPRL2000}
A. Cavagna, J. Garrahan, I. Giardina and D. Sherrington, Phys.
Rev. Lett. {\bf 85}, 5009 (2000).

\bibitem{HeimelCoolen}
J.A.F. Heimel and A.C.C. Coolen,  Phys. Rev. E {\bf 63}, 056121 (2001).

\bibitem{DeDominicis}
C. {De Dominicis}, Phys. Rev. B {\bf 18},  4913  (1978).

\bibitem{Coolen}
A.C.C. Coolen and J.A.F. Heimel, manuscript in preparation (2001).

\bibitem{SompZipp82}
H. Sompolinsky and A. Zippelius, Phys. Rev. B. {\bf 25},  6860
(1982).

\bibitem{HeimelDM}
A. De Martino and J.A.F. Heimel, manuscript in preparation (2001).

\end{references}
\end{document}